\documentclass[a4paper,11pt]{article}
\pdfoutput=1 

\usepackage{jheppub} 
\usepackage{amssymb}
\usepackage{amsmath}
\usepackage{color}
\usepackage{chngcntr}
\counterwithin{equation}{section}
\usepackage{graphicx}
\usepackage{bbm}
\usepackage{dsfont}
\usepackage{blindtext}
\usepackage{float}
\usepackage{caption}
\usepackage{subcaption}
\DeclareGraphicsExtensions{.png,.pdf}
\usepackage[T1]{fontenc} 
\usepackage{braket}
\usepackage{mathtools}
\usepackage{multicol}

\usepackage{color}

\DeclarePairedDelimiter\abs{\lvert}{\rvert}%
\makeatletter
\def\@fpheader{\relax}
\let\oldabs\abs
\def\abs{\@ifstar{\oldabs}{\oldabs*}}
\makeatother

\title{Anomaly induced quantum correction to charged black holes;  geometry and thermodynamics}

\author[a,b]{Jahed Abedi,}
\author[a,b]{Hessamaddin Arfaei}
\author[a]{Alek Bedroya,}
\author[a]{Aida Mehin-Rasuliani}
\author[a]{Milad Noorikuhani}
\author[a,c]{Kamran Salehi-Vaziri}

\affiliation[a]{Department of Physics, Sharif University of Technology, P.O. Box 11155-9161, Tehran, Iran}
\affiliation[b]{School of Particles and Accelerators, Research Institute for Fundamental Sciences, (IPM), P.O. Box 19395-5531, Tehran, Iran}
\affiliation[c]{Department of Physics and Astronomy, University of California, Riverside, California 92521, USA}

\emailAdd{jahed\_abedi@physics.sharif.ir}
\emailAdd{arfaei@sharif.ir}
\emailAdd{arfaei@ipm.ir}
\emailAdd{bedroya\_alek@physics.sharif.ir}
\emailAdd{mehinrasulian@physics.sharif.ir}
\emailAdd{noorikuhani\_milad@physics.sharif.ir}
\emailAdd{kamran.salehi@email.ucr.edu}

\abstract {We consider the corrections due to quantum fluctuations of fields on charged black holes induced from the energy-momentum trace anomaly. Although the number of horizons stays unchanged and their positions receive only finite corrections, the geometry,  thermodynamics and formation of RN black holes change seriously in particular for small ones. The entropy receives a logarithmic correction. The line $ Q=M $, separating naked singularities from physical solutions is corrected, putting a lower limit on the mass and an upper limit on the temperature of the black hole as a function of its charge. The modifications are highly significant in the cases of near-extremal and small black holes. We also show that for black holes with small mass can stay in thermal equilibrium without any constraint on the volume of the container. This result is in contrast to the large black holes that need a finite volume container for thermal equilibrium. The minimum of the mass lower limits occurs at zero charge, resulting in the extremal Schwarzschild black hole with a specific mass of the order of $M_{p}$ and zero temperature. This state which has only gravitational interaction will be the final stage of Hawking radiation.  Stability and lack of any interaction but gravitational, makes the extremal Schwarzschild black hole a serious candidate for dark matter particle.
}

\begin{document} 

\maketitle 

\section{Introduction}
\label{sec:intro}

Recent developments on the possible observation of Planck scale structure near the black hole horizon \cite{Abedi:2016hgu,Abedi:2017isz} has put more emphasis on the examination of the range of validity of classical treatment of gravity with quantum fields living on its background. Black holes have a long history of confronting us with a number of fundamental challenges of modern theoretical physics\cite{Hayward:2005gi,0264-9381-5-12-002,Abedi:2016hgu,Almheiri:2012rt,Maldacena:2013xja,Sekino:2008he,PrescodWeinstein:2009mp}. Some solved,  but still a number of them remain to be answered. Among them,  an important challenge is to find the limits of its classical description while the fields living in its background treated quantum mechanically \cite{Abedi:2015yga,Arfaei:2016dbh}. This also confronts us with the question of whether QFT effects results in new without invoking quantum gravity. This question was partially examined in \cite{Abedi:2015yga} in connection with the formation of singularity in the process of collapse of a massive shell of dust and later in \cite{Arfaei:2016dbh} for a ball of dust. It is shown that the formation of singularity is blocked due to positive pressure induced by zero-point quantum fluctuations of the fields living inside and outside the shell or the ball. In both cases, it was observed that due to fluctuations an inner horizon is formed similar to that of the charged Reisner-Nordstrom black hole which puts the radial direction back to a space like direction, leading to a halt of the collapse \cite{Abedi:2015yga}. Similar considerations were also applied to black holes in several articles \cite{Bambi:2013gva,Bambi:2013caa,Rovelli:2014cta,Bardeen:2014uaa,Mersini-Houghton:2014zka,Mersini-Houghton:2014cta,Modesto:2004xx,Torres:2014gta,Torres:2015aga,Chakraborty:2015nwa,Taves:2014laa,Vaz:2014era,Culetu:2014tqa,Hayward:2005gi,Kawai:2013mda,Kawai:2014afa,Barcelo:2014npa,Barcelo:2014cla,Kawai:2015uya,Allahbakhshi:2016gyj}. They either depend on arguments based on Hawking radiation, or quantum gravity effects. It is also shown that Hawking radiation is not strong enough to prevent the formation of neither the horizon nor the singularity. It  will only delay the process \cite{Paranjape:2009ib,Frolov:2014wja,Bardeen:2014uaa}.
Our main goal in this paper is to explore the quantum corrections induced by the conformal anomaly rising in the trace of energy-momentum tensor of the fields living in the background of interest. The correction is of the order of $\hslash$, comparable to Hawking radiation effects. In the present work, the Hawking radiation is considered only as a phenomenon occurring after the formation of the black hole. 
We would like to emphasize that there is a fundamental difference between our approach and those of others. In contrast to other attempts that try to modify the foundations of the physics of the black holes, the correction we consider is based on well established concepts and dynamics and hence is inevitable. If quantum fields exist, so do their fluctuations, and hence our observation on the deformation of the horizons of RN black holes, and in its limiting case the formation of the inner horizon for Schwarzschild black hole. Obstruction of singularity observed in \cite{Abedi:2015yga,Arfaei:2016dbh} neither requires any new postulate or assumption. 
Exact calculation of the contribution of the quantum fluctuation of the background fields to the zero point energy is a formidable task even when their interactions are ignored. In the flat background, it diverges and is set to zero by renormalization or other assumptions such as supersymmetry. The problem is highly nontrivial in a curved geometry. It has a strong dependence on the local conditions. It is well known that the trace of the energy-momentum tensor that is expected to be zero for massless fields is non-zero, hence it can not be put to zero or ignored. The nonzero value developed is a second order function of the background curvature. It is also proportional to $\hbar$. It is not significant when the curvature is not large, but it becomes important when the radius of curvature becomes comparable to the Planck's length. This condition occurs near the singularity, either for the cosmological solutions or for black holes, particularly small ones. It is worth to note that there are earlier attempts to study quantum-corrected black holes due to trace anomaly \cite{Nojiri:1998ph,Nojiri:1998ue,Nojiri:1999pm,Nojiri:2000ja,Elizalde:1999dw,Bytsenko:1998md}.
We will consider the energy-momentum tensor due to fluctuations as a perturbation. Of course, quantum fluctuations have been considered before in the form of Hawking radiation, but our consideration is about the correction to the static background geometry. It is worth noting that they are of the same order.

In this article we concentrate on the Reisner-Nordstrom black holes where the
coupled Einstein-Maxwell equations are:
\begin{align}
G_{\mu \nu} = T_{\mu \nu} + \braket{T_{\mu \nu}}\\
\nabla_{\mu} F^{\mu \nu} = \nabla_{\mu}*F^{\mu \nu} = 0
\end{align}
$\braket{T_{\mu \nu}}$ is the the expectation value of $T_{\mu\nu}$ due to quantum fluctuations in the background of black hole. Its trace in a general curved background has been calculated exactly as a quadratic function of the curvature \cite{PhysRevD.33.2840,Capper,Deser197645,Duff1977334,Duff:1993wm,PhysRevD.39.2993,Davies,Mottola:2010gp}. The anomaly is well studied but in order to be as self contained as possible we give a brief review in section \ref{Brief review of trace anomaly}. Moreover, we provide arguments to justify the use of our conclusions for Planckian black holes.

Section \ref{Geometric modifications} is devoted to developing a technical method which allows us to reconstruct all of the diagonal elements of the energy-momentum tensor from its trace by use of symmetries, in the order of $\hbar$. Then we obtain the modified metric putting back the quantum correction into Einstein-Maxwell equations and solve them to find the modified black hole solution. The modified black hole metric still have two horizons which are displaced by an amount of order of ${l_p}^2/M $. Interestingly this is the same region that quantum gravitational effects are expected to occur \cite{Abedi:2016hgu,Abedi:2017isz}.
We will see that the horizons of Reissner-Nordstrom black hole could be continuously
transformed into the horizons of modified Schwarzschild black hole of \cite{{Abedi:2015yga,Arfaei:2016dbh}} as the charge vanishes. In a loose language, one can say that the quantum corrections act like electrical charge for charge-less black holes by increasing the energy density near the center similar to the electric field, giving rise to two horizons.

Symmetries of RN  solutions are strong enough to give us sufficient information on the energy-momentum tensor, using the trace anomaly and find the corrections to the geometry and its effects. In our present considerations, the corrections to the diagonal terms are sufficient in contrast to the Hawking radiation which requires information on the off-diagonal terms $\braket{T_{t}^{r}}$. This quantum corrected geometry is very similar to the initial RN solution with the positions of the horizons modified with an amount of the O($l_{p}^{2}/M$). The internal horizon does not vanish when the charge goes to zero and approaches the internal horizon of the modified Scwartzchild BH.  Interestingly, the extremal solution of RN approaches the extremal  Schwarzschild solution \cite{Abedi:2015yga}. Moreover, we find that the region of physical solutions, i.e. solutions avoiding naked singularity,  $M\geq Q$ changes. In other words, the condition for the extremity of RN solution is modified. We distinguish two regions where we can apply a perturbative scheme to find the curve separating the physical solutions from solutions with a naked singularity. One is the region where the term dependent on the charge is large compared to the quantum correction. The other region, is where the charge is small and quantum correction has to be taken in the account first. This helps us to find the curve in two limiting cases.  We also find the geometric parameters for a generic case far from the limits. The extremal curve is also calculated by numerical methods.  As is seen in both perturbative and numerical approach, we recover the known results as charge and mass become large.

The next issue we consider in section \ref{Thermodynamics} is the corrections to the thermodynamics due to changes in the geometry. The main change takes place for the small black holes and in particular for small charges. The entropy receives a logarithmic correction as the signal of $l_{p}^{2}/M$ correction to the position of the outer horizon which also comes from Planck proper distance away from the horizon\cite{Abedi:2016hgu,Abedi:2017isz}. This has its roots in higher order curvature terms appearing in the equations. Such logarithmic terms also appear in other approaches \cite{Das:2001ic}. It is known that RN black holes with fixed charge pass through a maximum temperature before reaching the extremal limit.  We find that there is an upper bound to this temperature which occurs when the charge goes to zero, setting an absolute maximum temperature for black holes due to quantum corrections. The specific heat changes sign at this temperature, indicating a phase transition. For any given charge we find a minimum mass with zero temperature which approaches the absolute non-zero minimum mass when the charge goes to zero; a cold remnant at the Planck scale, solely with gravitational effects, as the final state surviving at the end the Hawking radiation.

The chemical potential also receives correction which confirms the observations made.

At small scales, the angular momentum barrier becomes very large and comparable to Planck mass allowing only very large quanta to escape as the result of Hawking radiation, with very low probability making black holes with small charge and mass as metastable states.

At section \ref{Applications} we consider two examples of applications of our results; equilibrium of black hole in thermal bath and collapse of a charged dust shell. We find that the small black holes inside a container with positive heat capacity stay stable without any constraint on the volume. While for the large black holes they cannot remain stable unless we put an upper bound on the volume of the container. We also show that for a collapsing charged shell it comes to a halt before reaching the singularity and after passing the internal horizon.

In the appendices, we will give details of some mathematical analysis on the number of horizons,  and non-analyticity of the extremal curve in Q-M space.

\section{Brief review of trace anomaly\label{Brief review of trace anomaly}}
The contribution of quantum fluctuations of fields to the background energy, momentum and other physical quantities in the vacuum state of a theory in flat space-time is usually set to zero by the normal ordering of the operators. This is a benign and harmless procedure in a field theory living in a flat background, but when gravity is concerned the situation is dramatically different since it couples to the energy-momentum tensor living in space-time independent of their origin. These quantum sourced fluctuations are usually small so that they can be ignored in many problems. Even though induced energy momentum tensor is of the order of $\hbar$, it is quadratic in curvature. That is why the problem becomes important when the background deviates significantly from flat and the curvature is large. Calculation of the full tensor in a general background is a formidable task, but its trace is at hand. We have observed that in certain cases with sufficient symmetry, the relevant elements of the tensor is obtained from the trace.

Classically we expect that in presence of conformal invariance, where there is no natural scale, the trace of the energy-momentum tensor should vanish. Due to the divergence of the product of two fields, when their arguments coincide, this classical expectation is frustrated and one gets a finite result for $\braket{T^\mu_\mu}$, which breaks the scale invariance. To set an example, consider the electromagnetic field, $A_\mu$, in a curved background given by a metric $g_{\mu\nu}$.
The Einstein Maxwell action for the system is,
\begin{align}
S=-\frac{1}{4\pi G}\int\sqrt{-g}[\mathcal{R}+\frac{1}{4}F_{\mu\nu}F^{\mu\nu}]
\end{align}
with the energy momentum tensor,
\begin{align}
T_{\mu\nu}(x)=F_{\mu\alpha}(x) F^{\alpha}_{\nu}(x)-\frac{1}{4}g_{\mu\nu}(x)F(x)^2
\end{align}

When we consider it as a quantum operator, involving  the product of two fields at the same point, it gives rise to an infinity. In a flat background it can be harmlessly removed and set to   zero.  In contrast in a curved  manifold  the infinity rising from operator products becomes position dependent. After removing an infinite constant as in the flat case, we are left with  a finite value for the trace $\braket{T^\mu_\mu}$. This trace which appears only after considering the quantum effects, is known as trace anomaly first calculated by Capper and Duff in 1973 \cite{Capper,PhysRevD.33.2840,Deser197645,Duff1977334,Duff:1993wm,PhysRevD.39.2993,Davies,Mottola:2010gp} and is given as,

\begin{equation}
\label{eq:trace_T}
\begin{split}
\braket{T_{\rho}^{\rho}} = \frac{\hbar}{32\pi} \{ (c_A+c_A^{\prime}) (\mathcal{F}
-\frac{2}{3} \Box \mathcal{R}) - c_A^{\prime} \mathcal{E} + c_A^{\prime\prime} \Box \mathcal{R} \}
\end{split}
\end{equation}
where

\begin{equation}
\label{eq:def_F}
\begin{split}
\mathcal{F}=\mathcal{C}_{\mu{}\nu{}\rho{}\sigma{}}\mathcal{C}^{\mu{}\nu{}\rho{}\sigma{}}=\mathcal{R}_{\mu{}\nu{}\rho{}\sigma{}}\mathcal{R}^{\mu{}\nu{}\rho{}\sigma{}}-2\mathcal{R}_{\mu{}\nu{}}\mathcal{R}^{\mu{}\nu{}}+\frac{\mathcal{R}^2}{3}
\end{split}
\end{equation}
and

\begin{equation}
\label{eq:def_E}
\begin{split}
\mathcal{E}=\mathcal{R}_{\mu{}\nu{}\rho{}\sigma{}}\mathcal{R}^{\mu{}\nu{}\rho{}\sigma{}}-4\mathcal{R}_{\mu\nu}\mathcal{R}^{\mu{}\nu{}}+\mathcal{R}^2
\end{split}
\end{equation}
in which $\mathcal{R_{\mu \nu \rho \sigma}}$, $\mathcal{R}_{\mu \nu}$, $\mathcal{R}$, and $\mathcal{C}_{\mu \nu \rho \sigma}$  stand for the Riemann curvature tensor, Ricci curvature tensor, Ricci scalar, and Weyl conformal tensor respectively. $c_A$ and $c_A^{\prime}$ are constants representing the relative contribution of number of massless fields living in the background,

\begin{equation}
\label{eq:def_c}
\begin{split}
c_A=\frac{1}{90\pi{}}(n_0+\frac{7}{4}n_{\frac{1}{2}}^M+\frac{7}{2}n_{\frac{1}{2}}^D-13n_1+212n_2)
\end{split}
\end{equation}
and

\begin{equation}
\label{eq:def_cprime}
\begin{split}
c_A^{'}=\frac{1}{90\pi{}}(\frac{1}{2}n_0+\frac{11}{4}n_{\frac{1}{2}}^M+\frac{11}{2}n_{\frac{1}{2}}^D+31n_1+243n_2)
\end{split}
\end{equation}
where $n_s$ is the number of spin s  particles. The superscripts $M$ and $D$   stands respectively for Majorana and Dirac spinors. The expressions above are valid for the massless free fields coupled conformally to the background.  Although massive fields give similar contributions, they are negligible compared to that of massless fields.

As mentioned earlier obtaining the full tensor from the anomaly is certainly not easy, however when the background enjoys symmetry, the conservation of energy-momentum tensor, $\nabla_\mu T^{\mu\nu}=0$, is a powerful constraint allowing us to go beyond the trace.    In the case of charged static black hole, symmetries are sufficient to allow us to complete the task and find,  if not the full energy-momentum tensor, but what we need i.e. the diagonal elements,  from its trace. The symmetries consist of time translation, rotational invariance and the emergent symmetry called radial boost. 

The expression \ref{eq:trace_T} is obtained based on one-loop calculations and so is exact for the free fields. One might wonder whether this expression is also reliable for interacting theories or higher loops would spoil the conclusions. Following, we address this natural question and using different arguments for backgrounds of low and high curvatures, we show the expression \ref{eq:trace_T} is reliable for both cases.

It is well known that in four dimensions, the conformal anomaly has the following structure, \cite{Hawking:2000bb,Komargodski:2011vj,Netto:2015cba},
\begin{equation}
\label{conformalstructure}
\braket{T_{\rho}^{\rho}} = c \mathcal{F} - a \mathcal{E} + d \Box \mathcal{R}
\end{equation}
where the coefficients $a$, $c$ and $d$ depend on the underlying QFT. Unfortunately, the coefficient $d$ is ill-defined since it depends on the regularization scheme, however it does not appear in the first order correction to the RN metric. Therefore, we can neglect $d$ in the first order perturbative analysis. Note that the higher order loop calculations change the coefficients $b$ and $a$ but do not affect the structure \ref{conformalstructure}.

Corrections to $c$ and $a$ from higher loops are functions of coupling constants of the theory that flow according to the RG equations. Hence, $a$ and $c$ have different values depending on the scale of our interest.

The energy scale required for the RG flow could be obtained from the energy density $\braket{T^{t}_{t}}$. In the case of our interest, high energy density implies highly curved background as well. Furthermore, for such cases, the energy density is of the same order as $\sqrt{\mathcal{R}}$. Therefore, $\sqrt{\mathcal{R}}$ can be used as energy scale for RG flow. The physical reason behind this is that when curvature surpasses other mass scales of the field theory it becomes the only natural available energy scale. The relation between curvature and the RG energy scale implies that for highly curved background we would face the UV completion of the field theory. A number of field theories have UV limits which are either free or with small couplings \cite{Komargodski:2011vj,Komargodski:2011xv}, such as asymptotically free and $\phi^4$ theories. For theories of free UV limit, which includes a broad range of interesting theories, the expression \ref{eq:trace_T} becomes exact. All in all, this means that the expression \ref{eq:trace_T} becomes more reliable for large curvatures (Planckian black holes). 

Before the present work, the conformal anomaly was used for investigation of scalar-free inflation. In that work and other articles regarding the effects of the conformal anomaly in GR the higher-loops contributions has been neglected (Hawking et al \cite{Hawking:2000bb}) or glossed over very briefly \cite{Netto:2015cba,Hawking:2000bb}. This is justified since their consideration is in either the early universe or Planck scale regimes where large curvature moves us into the deep UV limit of the field theory.

As curvature decreases, such as moving away from the center of the BH, the energy density decreases as well. This implies that the theory flows towards its IR limit for low curvature backgrounds. Along the RG flow, the coupling constants increase and the impact of interaction terms becomes more significant. As a result, the higher order loop calculations should be considered and the values of $a$ and $c$ would be corrected. It has been shown that the changes in these coefficients are small compared to their UV values \cite{Komargodski:2011vj,Komargodski:2011xv} and so could be neglected.

In support of reliability of the mentioned argument for the low-energy limit, we remark that for backgrounds of small curvature, the corrections from conformal anomaly drop to zero much faster than the classical terms. Therefore, the UV terms that are larger than the change due to the RG flow, are suppressed by the classical energy-momentum tensor. This means that in the IR limit, we can safely consider the UV expressions for the first order perturbative analysis.

In the following, for our general conclusions, we do not adhere to any particular values for $c$ and $a$, however, for concrete examples the one-loop results for $a$ and $c$ are used.

\section{Geometric modifications\label{Geometric modifications}}
In this section, we first derive the diagonal elements of the energy-momentum tensor induced by the anomaly in the RN background and then solve the new Einstein-Maxwell equations to find the modified metric.  We continue with the study of the properties of the new metric and find the new horizons and the modified extremal conditions.
In order to simplify the calculations, we set $G$, $c$, and $\epsilon_0$ equal 1, i.e. giving mass and charge in Planck's units.

\subsection{Diagonal elements of the energy momentum tensor}
\label{sec: diagonal T}

The trace of the effective energy-momentum tensor resulted by the back reaction of the quantum fields, is given by \eqref{eq:trace_T}. We try to extract the necessary elements of energy-momentum tensor using the trace (conformal) anomaly. 

Since we are looking for a spherical symmetric static charged solution,  $\braket{T^\mu_\nu}$ is diagonal and $\braket{T^\theta_\theta}=\braket{T^\phi_\phi}$. On the other hand, since Riemann tensor remains unchanged under radial Lorentz boost, $G^\mu_\nu$ will not change either. Hence $\braket{T_{\mu}^{\nu}}$, which is proportional to $G^\mu_\nu$, would also remain unchanged. By writing radial Lorentz transformation in details, one can find that any diagonal second rank tensor, say $\braket{T_\mu^\nu}$, would remain unchanged if and only if $\braket{T^r_r}=\braket{T^t_t}$.
Furthermore, Dymnikova and Bardeen also indicated that zero point energy \cite{Dymnikova:2001fb, Dymnikova:2003vt} including the conformal anomaly part \cite{Bardeen:2014uaa} (see also \cite{PhysRevD.56.2180, Ansoldi:2008jw}) should respect all of the symmetries of the Schwarzschild (or Reissner Nordstrom) geometry. So in this case, the symmetry of the anomalous source is reduced from the full Lorentz group in flat space-time to the Lorentz boosts in the radial direction. Therefore, we conclude that energy-momentum tensor due to fluctuations of the fields in a static spherically symmetric background should take the following form;

\begin{align}
\braket{T_\mu^\nu}=\left(\begin{array}{
cc}
\begin{array}{
cc}
\kappa(r) I_2
\end{array} & \begin{array}{
cc}
0
\end{array} \\
\begin{array}{
cc}
0
\end{array} & \begin{array}{
cc}
\lambda(r) I_2
\end{array}
\end{array}\right)
\end{align}

Another constraint on $\braket{T_\mu^\nu}$ is obtained from the fact that it is divergence free;
\begin{align}
\frac{d\kappa(r)}{dr}+\frac{2\kappa(r)}{r}-\frac{2\lambda(r)}{r}=0
\label{div-free}
\end{align}
This  equation puts a strong condition  on the general form of $\braket{T_{\mu}^{\nu}}$. In order to see this, let us consider the case in which $\braket{T_{\mu}^{\nu}}$ is proportional to $r^{-p}$. In this case, $\frac{d\kappa(r)}{dr}=-\frac{p}{r}\kappa(r)$ and so the equation \eqref{div-free} turns into the following form,
\begin{align}
\label{fraction}
\frac{\lambda(r)}{\kappa(r)}=-\frac{p}{2}+1
\end{align}
If we expand $\braket{T_\mu^\nu}$ in terms of inverse powers of $r$, the above equation implies that,
\begin{align}
\label{Tform}
\braket{T_\mu^\nu}=\sum\limits_{p \in I}T_p
\left(\begin{array}{
cc}
\begin{array}{
cc}
I_2
\end{array} & \begin{array}{
cc}
0
\end{array} \\
\begin{array}{
cc}
0
\end{array} & \begin{array}{
cc}
(-\frac{p}{2}+1)I_2
\end{array}
\end{array}\right)
r^{-p}
\end{align}

This is a powerful constraint on the form of $\braket{T_\mu^\nu}$ such that we can uniquely find it from its trace. Based on the equation \eqref{Tform}, trace of $\braket{T_\mu^\nu}$ can be written as:
\begin{align}
\label{traceformula}
\braket{T_\mu^\mu}=\sum\limits_{p\in I} (4-p)T_p r^{-p}
\end{align}
Hence one can find coefficients $T_p$ and the energy momentum tensor $\braket{T_\mu^\nu}$ by expanding the trace in terms of inverse powers $r$. It is worth  remarking  that the equation \eqref{traceformula} implies that gravitational part of the trace of energy momentum tensor 
will be lacking the  $r^{-4}$ term. It also implies  that the only chance for $\braket{T_{\mu}^{\nu}}$ to become traceless is to change as $r^{-4}$.

We are looking for a static spherical solution to the modified field equations given as,
\begin{align}
\mathcal{R}_{\mu\nu}-\frac{1}{2}g_{\mu\nu}\mathcal{R}=8\pi\braket{T_{\mu\nu}}+T_{\mu\nu}
\label{modifiedfieldeq}
\end{align}
with mass $M$ and electric charge $Q$. We  remark that $T_{\mu\nu}$ is the classical electromagnetic energy momentum tensor and $\braket{T_{\mu\nu}}$ is the backreaction of the quantum fields derived from its trace. The energy momentum tensor depends on quantum variables $c_A$, $c'_A$, and $c''_A$ as well as geometrical variables $R_{\mu\nu\rho\sigma}$. By changing quantum variables $c_A$, $c'_A$, and $c''_A$, the  solution would also change. The right hand side of the equation \eqref{modifiedfieldeq} is at least second order in terms of $l_p$, so the lowest order component of the metric should satisfy the Einstein equation without the quantum contribution to the energy momentum tensor.   Hence the zeroth order component is given by the classical Reissner Nordstrom solution.
\begin{equation}
\label{eq:classic_RN}
\begin{split}
ds^2 = (1-\frac{2M}{r}+\frac{Q^2}{r^2}) dt^2 - ( 1 - \frac{2M}{r}+\frac{Q^2}{r^2})^{-1} dr^2 - r^2 d\Omega
\end{split}
\end{equation}
Substituting the zeroth component of metric into the right hand side of equation \eqref{modifiedfieldeq}, one can find the next order of the left hand side up to the lowest order in terms of $l_p^{2}$,
\begin{equation}
\label{eq:trace anomaly of T}
\begin{split}
\braket{T_{\mu}^{\nu}}=-\frac{3c_Al_p^2}{4\pi{}}\left(\begin{array}{
cc}
\begin{array}{
cc}
I_2
\end{array} & \begin{array}{
cc}
0
\end{array} \\
\begin{array}{
cc}
0
\end{array} & \begin{array}{
cc}
-2 I_2
\end{array}
\end{array}\right)\frac{M^2}{r^6}\ +\frac{c_Al_p^2}{2\pi{}}\left(\begin{array}{
cc}
\begin{array}{
cc}
2I_2
\end{array} & \begin{array}{
cc}
0
\end{array} \\
\begin{array}{
cc}
0
\end{array} & \begin{array}{
cc}
-5 I_2
\end{array}
\end{array}\right)\frac{MQ^2}{r^7} 
\\
-\frac{(6c_A+c_A^{\prime})l_p^2}{16\pi{}}\left(\begin{array}{
cc}
\begin{array}{
cc}
I_2
\end{array} & \begin{array}{
cc}
0
\end{array} \\
\begin{array}{
cc}
0
\end{array} & \begin{array}{
cc}
-3I_2
\end{array}
\end{array}\right)\frac{Q^4}{r^8}
\end{split}
\end{equation}
where $I_2$ is $2\times2$ unit tensor. By inserting the above results in (\ref{modifiedfieldeq}) we obtain the field equations for the metric up to first order quantum corrections.

\subsection{Modified metric}
\label{sec: new metric}

Now we proceed to solve the Einstein-Maxwell equations for the energy-momentum tensor induced by the anomaly. The energy momentum tensor is quadratic in curvature tensor, hence the equation has transformed to a modified Einstein-Maxwell equation where the second power of the curvature has been added to it. Although the solution to such highly nonlinear equation is rather impossible, we will treat the extra term as a perturbation. The extra term is of the order of  $ \hslash$. On the other hand,  we are looking for the metric corrections to the same order, therefore we use the energy-momentum tensor in the background of the original RN black holes (\ref{eq:trace anomaly of T}).

Symmetry requirements imply that,
\begin{equation}
\label{eq:general spherical metric}
\begin{split}
ds^2=f(r)dt^2-g(r)dr^2-r^2(d{\theta{}}^2+\sin^2{\theta{}}\
d{\phi{}}^2)
\end{split}
\end{equation}

The condition $\braket{T^t_t}=\braket{T^r_r}$ implies that $f(r)=g(r)^{-1} $, which is  respected also in higher orders of iterations. 
 The $t t$ component  of the Einstein tensor could be written as:

\begin{equation}
\label{eq:gtt}
\begin{split}
G_t^t=\frac{1}{r}\frac{d}{dr}f-\frac{1}{r^2}f+\frac{1}{r^2}
\end{split}
\end{equation}
By substituting \eqref{eq:trace anomaly of T} and \eqref{eq:gtt} in the Einstein's equation, we have,

\begin{equation}
\label{eq:solving1}
\begin{split}
\frac{1}{r}\frac{d}{dr}f-\frac{1}{r^2}f+\frac{1}{r^2}=-\frac{6c_Al_p^2M^2}{4}\frac{1}{r^6}+8c_AMQ^2 \frac{1}{r^7}-\frac{\left(6c_A+c_A^{\prime}\right)l_p^2Q^4}{2}\frac{1}{r^8}
\end{split}
\end{equation}
which results in,

\begin{equation}
\label{eq.3.12}
\begin{split}
f(r)=1-\frac{2M}{r}+\frac{Q^2}{r^2}+c_Al_p^2\left(\frac{2M^2}{r^4}-\frac{2MQ^2}{r^5}+\frac{\alpha Q^4}{r^6}\right)
\end{split}
\end{equation}
where $\alpha = \frac{3}{5}+\frac{c_A^{\prime}}{10c_A}$. 
For a generic $Q$ and $M$, one expects the new horizons to receive small corrections proportional to $c_{A} l_{p}^{2}/M$. There are two regions that need special care; the region near $Q=0$, and the region near $Q=M$. In the region near $Q=0$, we cannot take the quantum correction as a perturbation to RN solution. In this case, the contribution of the electromagnetic energy-momentum tensor is considered as a perturbative term over the quantum corrected Schwarzschild solution \cite{Abedi:2015yga}. On the other hand at $Q=M$, the first-order derivation of the classical RN solution vanishes and first order perturbation of quantum corrected RN is not reliable. In this case, we see the boundary $Q=M$ which separates the physical solution from solution with naked singularity changes and receives correction. In the following, we first find this boundary in the two cases of small charge and large charge and then proceed to obtain the changes in horizons for a generic case. The boundary of the physical solutions and naked singularity represents the extremal solutions, i.e. solutions with a double horizon where the inner and outer horizon coalesce. At a double horizon (extremal) both $f(r;Q,M)$ and its first derivatives vanish,

\begin{equation}
f(\tilde{r}_{ext};Q,M)=f'(\tilde{r}_{ext};Q,M)=0. \label{eq.3.13}
\end{equation}
We take M as given and find Q and $\tilde{r}_{\pm}$ as functions of M.

For classical RN solution without quantum corrections $f$ approaches $f_{0}(r,M,Q)=1-\frac{2M}{r}+\frac{Q^2}{r^2}$, resulting in  the extremal values of Q and $r_{ext}$ ,
\begin{equation}
r_{ext}=Q=M
\end{equation}

Where the quantum  corrections are considered , we assume   the extremal values to get a small correction as following, 
\begin{equation}
\tilde{r}_{ext}=r_{ext}(M)+c_{A}l_{p}^{2}\delta r_{ext}
\end{equation}
\begin{equation}
\tilde{Q}=Q(M)+c_{A} l_{p}^{2} \delta Q
\end{equation}

Now (\ref{eq.3.13}) with $f(r,t)=f_0(r,t)+f^1(r,t)$
takes the form,
\begin{eqnarray}
  \begin{cases}
               f_{0}(r,Q,M) + c_{A}l_{p}^{2} f_{1}(r,Q,M)=0\\
               f_{0}'(r,Q,M) + c_{A}l_{p}^{2} f_{1}'(r,Q,M)=0
            \end{cases}
\end{eqnarray}

Expanding the equation around classical solutions and noting that $f_{0}$ and $f'_{0}$ vanish for Q=M, we can find Q and $r_{\pm}$.
\begin{equation}
\tilde{Q}=M-c_{A} l_{p}^{2} \frac{\alpha}{2M}
\end{equation}
\begin{equation}
\tilde{r}_{ext}=M+c_{A} l_{p}^{2} \frac{2 \alpha -1}{M}
\end{equation}

Note that the line Q=M appears as the asymptote for the corrected extremality condition at large $M$. One can loosely interpret that $\frac{c_{A}l_{p}^{2}}{2M}$ acting  as extra electric charge over the true electrical charge $Q$.  This interpretation sheds light on the acquiring of a small internal horizon of Schwarzschild as the result of quantum correction.

It is also worth noting that the extremal radius slightly changes compared to the classical case. Again for large mass RN black holes, the correction vanishes.

\begin{figure}
\centering
\includegraphics[width=1\linewidth]{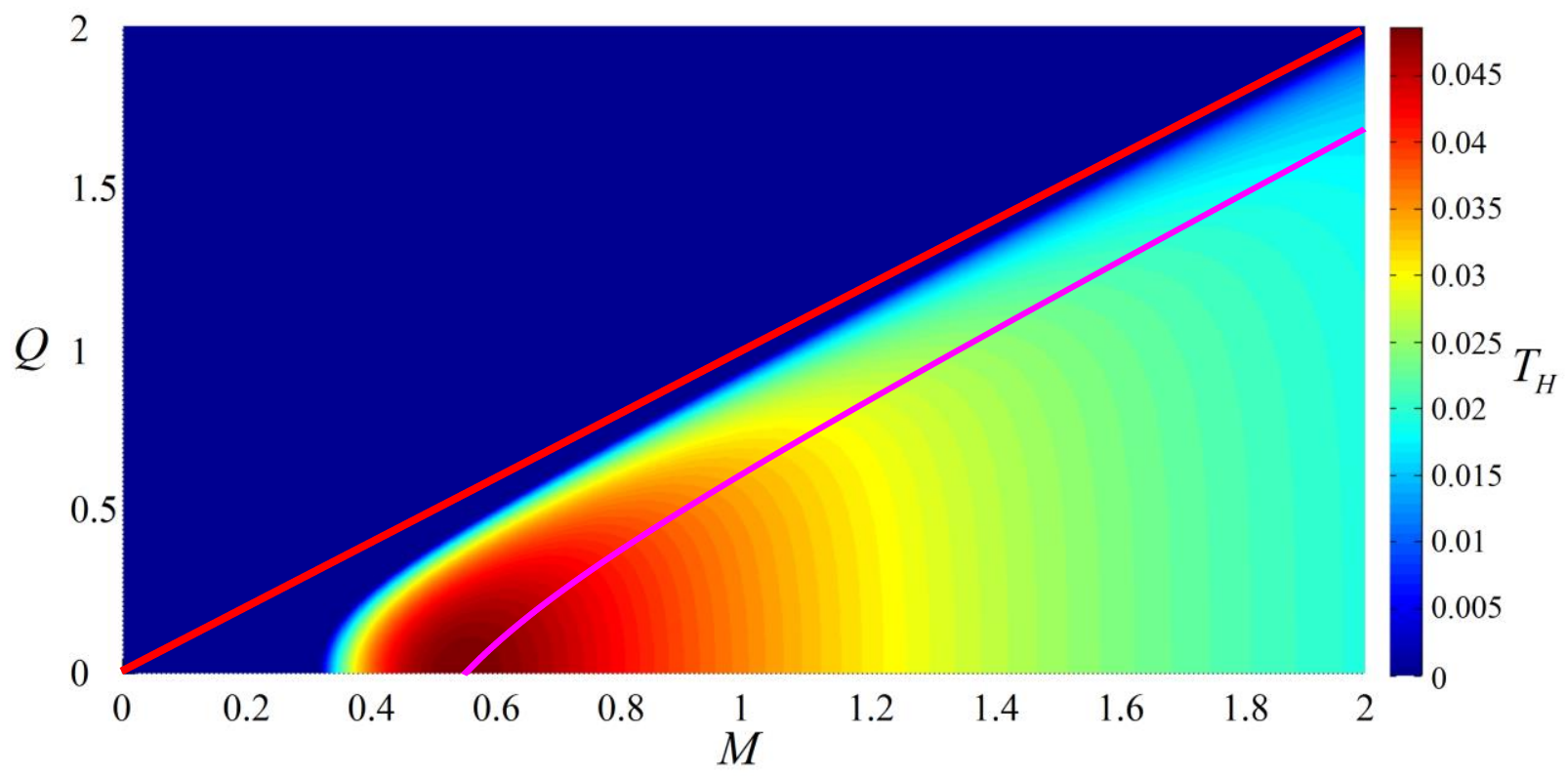}
\caption{Quantum corrected temperature for Reissner Nordstrom black hole given in Eq. (\ref{eq.4.1}). The straight red line is the boundary of extremal for classical RN black hole. The blue region is representing the region of naked singularity for quantum corrected RN black hole. The purple curve is representing the points with maximum temperature.}
\label{Temperature}
\end{figure}

The other region in need of special attention is where Q is small; i.e perturbing the Schwarzschild black hole by adding small chrges to it. We begin with quantum corrected Schwarzschild solution of \cite{Abedi:2015yga} and take $Q^{2}$ as expansion parameter for the perturbation calculations. We can decompose $f(r,M,Q)$,
\begin{equation}
f(r,M,Q)=g_{0}(r,M,Q)+Q^{2}g_{1}(r,M,Q)
\end{equation}
where,
\begin{equation}
g_{0}=(1-\frac{2M}{r}+c_{A}l_{p}^{2}\frac{2M^{2}}{r^{4}}), \ \ g_{1}=\frac{1}{r^{2}}(1-c_{A}l_{p}^{2}\frac{2M}{r^{3}}). \label{eq.3.21}
\end{equation}
We expand around extremal radius of the quantum corrected Schwarzschild black hole where $g_{0}$ and $g'_{0}$ both vanish.
We take $c_{A}$ as given since it depends only on the number of types of the fields. The extremal values for mass and radius turn out to be,
\begin{equation}
\tilde{r}_{ext}=\frac{3}{2}M_{ext}=\sqrt{\frac{8}{3}c_{A}}l_{p}
\end{equation}
The new values as solutions to $f=f'=0$ can easily found to be,
\begin{equation}
\tilde{M}_{ext}=\sqrt{\frac{32}{27}c_{A}}l_{p}+ \sqrt{\frac{3}{32}} \frac{Q^{2}}{\sqrt{c_{A}} l_{p}}
\end{equation}
and
\begin{equation}
\tilde{r}_{ext}=\sqrt{\frac{8}{3}c_{A}}l_{p}+\frac{\sqrt{3}}{\sqrt{512}} \frac{Q^{2}}{\sqrt{c_{A}}l_{p}}
\end{equation}
In this part we have calculated $M_{ext}$ and $r_{ext}$ in terms of a given $Q$.
Hence the line Q=M changes to asymptote of the curve starting from $M_{ext}=\sqrt{\frac{32}{27}c_{A}}l_{p}$ at Q=0 with infinite slope gradually approaching the old boundary (See figure \ref{Temperature}). The range of $(Q,M)$ above this curve corresponds to naked singularity.

\begin{figure}
\centering
\includegraphics[width=0.4\linewidth]{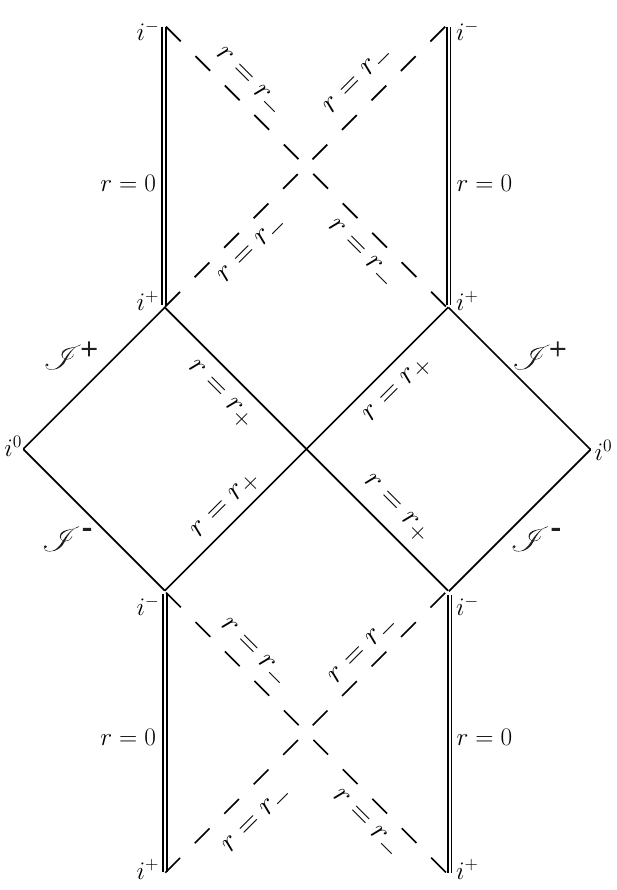}
\caption{Penrose diagram for the static line element in (\ref{eq:general spherical metric}).}
\label{penrose}
\end{figure}

Now that we have found the range of physical solutions, we proceed to obtain the corrections to the horizons for a generic case Q, and M far from Q=M and Q=0.
We have to solve only  the first equations in (\ref{eq.3.13}) , i.e. $f(\tilde{r}_{\pm},Q,M)=0$  keeping $Q$ and $M$ fixed. The result is, 
\begin{equation}
\tilde{r}_{\pm}=r_{\pm} - \frac{c_{A}l_{p}^{2}}{M \eta_{\pm}^{2}} \left(  1+ \frac{\alpha}{2} \frac{ (\eta_{\pm} - 2)^{2} }{\eta_{\pm}-1}  \right)
\end{equation}
with $\eta_{\pm}=1 \pm \sqrt{1-(Q/M)^{2}}$. We note that we have assumed $QM>>c_{A}l_{p}^{2}$. Penrose diagram of this black hole which is shown in Fig. \ref{penrose}, is the same as Reissner Nordstrom black hole. Following the same steps, but this time for the new perturbation regime $QM<<c_{A}l_{p}^{2}$ we have,
\begin{align}
\tilde r_+&=2M-\frac{c_Al_P^2}{4M}-\frac{Q^2}{2M}\nonumber\\
\tilde r_-&=c_A^\frac{1}{3}l_P^\frac{2}{3}M^\frac{1}{3}-\frac{6Q^2}{M}
\end{align}

We remark that quantum corrections, similar to the Maxwell field effectively increase the energy density of black hole interior, which results in decreasing and then increasing the radius of the inner horizon and decreasing the outer one. This justifies the fact that even after quantum corrections we still find two horizons, although the degree of the equation changes from two to six. 
The $Q=0$  case inherits this from RN black hole and therefore we expect at most two horizons even in this case. The mathematical details of the proof are given in the appendix \ref{A1}.

It is worth noting that in terms of  $c_{A}$ the behavior of the curve separating solutions with naked singularity from the physical ones is not analytic near Q=0, but approaches to analytic form as we move away from it. It becomes dominantly linear at large values of Q.
As a result, the inner horizon in this limit does not behave analytically either (see appendix \ref{A2}).
Similar to the inner horizon of standard RN black hole which has its root in nonzero energy-momentum tensor of the Maxwell fields, the inner horizon of Q=0 results from nonzero energy-momentum tensor due to vacuum fluctuations. Same phenomenon contributes to enlarge inner horizon to the RN solution and making the outer one smaller.

Although the mass and size that we have found for the extremal Schwarzschild black holes are not exact and may receive correction under further iterations, the general picture that has emerged, i.e. the existence of a stable, charge-less,  zero temperature,  Planck size black hole with only gravitational interaction survives. We give a general argument in support of this picture. The argument depends on the assumption that flat Minkowski space-time is stable. This means that any fluctuation in the Minkowski space will add to its energy. Therefore the energy of a region with fluctuation in non-zero curvature is positive. Combining this point with the fact that the trace of the energy-momentum tensor is quadratic in curvature, we can conclude that the complete energy-momentum tensor due to field fluctuations must be at least quadratic in curvature. Therefore the regions with a high curvature such as the interior of a black hole contain large energies. This large energy acts the same way the positive large energy density of a charged black hole acts near the center leads to an internal horizon. If the energy were not positive, then it would be energetically more desirable for the space to form seeds of negative energy and shed the positive part to infinity. In other words, the  flat space would become unstable and the curvature would grow indefinitely. So independent of the numerical details we expect an inner horizon to form. As the mass of the black hole decreases the new inner horizon and the outer one will eventually meet and form a black hole with a double horizon, the extremal black hole.

Now that we have studied the geometric aspects of the quantum corrections to RN solutions, we will take a different direction to study the effects of the fluctuations on its thermal properties.

 \section{Thermodynamics\label{Thermodynamics}}

In this section, we obtain the correction to the thermodynamic quantities such as temperature, chemical potential and entropy due to quantum corrections of  the geometry.  We will see that the corrections become substantial especially when Q becomes small.
First, we study the general case in \ref{sec:nonzero thermal} and then we restrict our considerations to zero charge limit in \ref{Thermodynamics of special case of Q=0}.

\subsection{Thermodynamic properties for non-zero charge}
\label{sec:nonzero thermal}

The temperature can be obtained from Hawking radiation assuming the perturbation around classical RN temperature $T_{H}^{RN}$,
\begin{equation}
T_{H}=T_{H}^{RN} - c_{A}l_{p}^{2} \frac{\eta_{+}-1}{\pi M^{3} \eta_{+}^{5}} \left( 1+\frac{3\alpha}{4} \frac{(\eta_{+}-2)^2 (\eta_{+}-2/3)}{(\eta_{+}-1)^{2}} \right) \label{eq.4.1}
\end{equation}
where $T_{H}^{RN}= (\eta_{+}-1) \left/ \left( 2\pi M \eta_{+}^{2} \right) \right.$.
Note that the line $Q=M$, or $\eta_{+}=1$   no longer represents the $T_{H}=0$ state; the first term of the correction vanishes at this line but the second term diverges negatively. This is another indication that the classically extremal case has moved to a nonphysical region with naked singularity as the result of quantum correction.  The $T_{H}=0$ curve deviates from its asymptote as the mass moves down from large values. Hence the deviation from the classical extremality condition becomes larger for small mass and charge, i.e. the charge to mass ratio becomes smaller than the classical case to the extent that the zero charge limit acquires a finite mass. The finite mass of the zero charge limit of the extremal case is of the order of Planck's mass. This mass sets a lower limit to the mass of all black holes. One can see from Fig. \ref{Temperature} that for a given charge $Q$ there is a minimum mass to the black hole, which is strictly larger than $Q$. When $Q$ decreases the minimum mass also decreases and approaches the absolute minimum for Schwarzschild black hole when the charge vanishes. On the other hand, there is an upper limit on the temperature for fixed charge as is well known for RN black holes. This upper limit increases as charge decreases and attains its strict maximum at zero charge. 

\begin{figure}[t]
\begin{multicols}{2}
    \includegraphics[width=\linewidth]{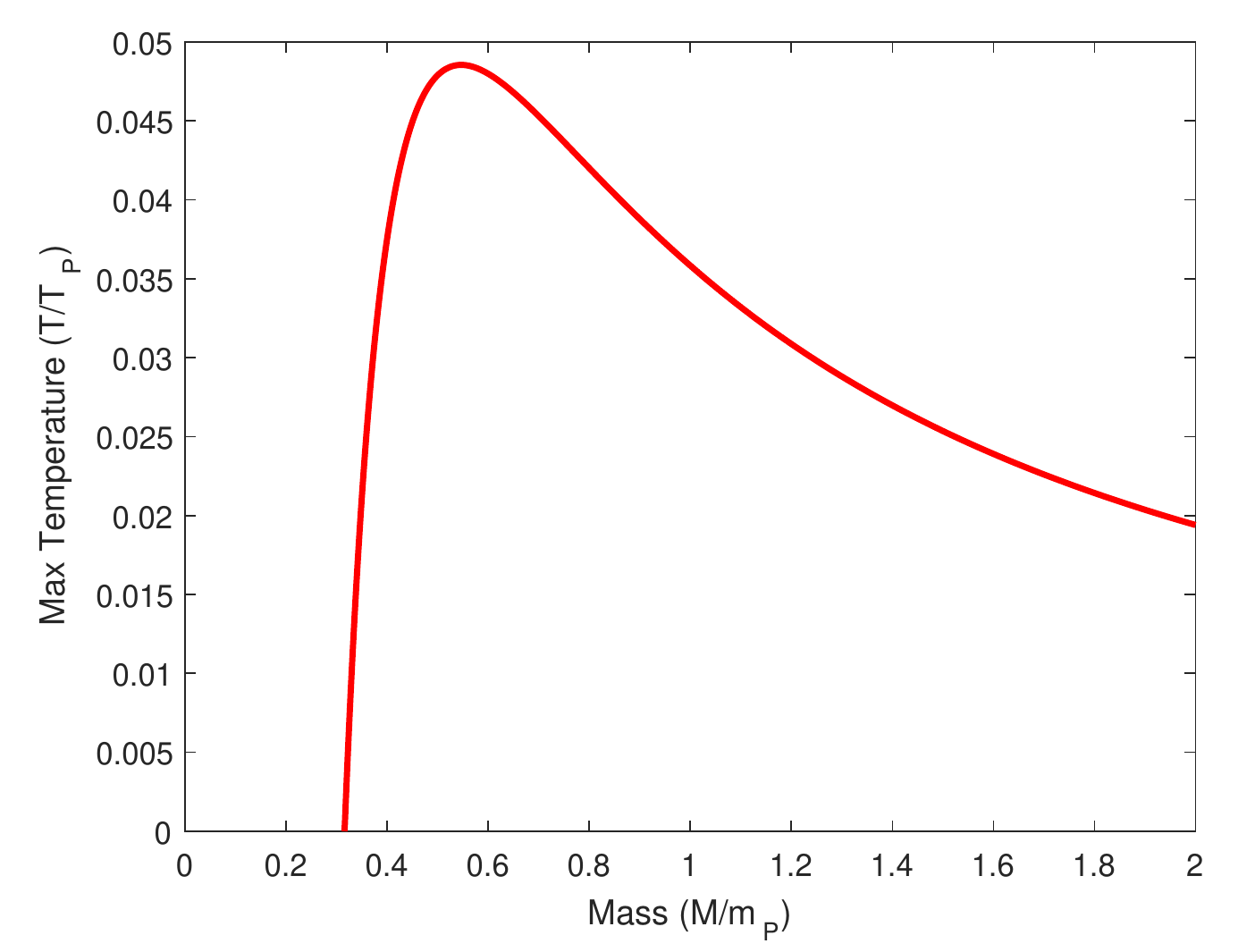}\par
    \includegraphics[width=\linewidth]{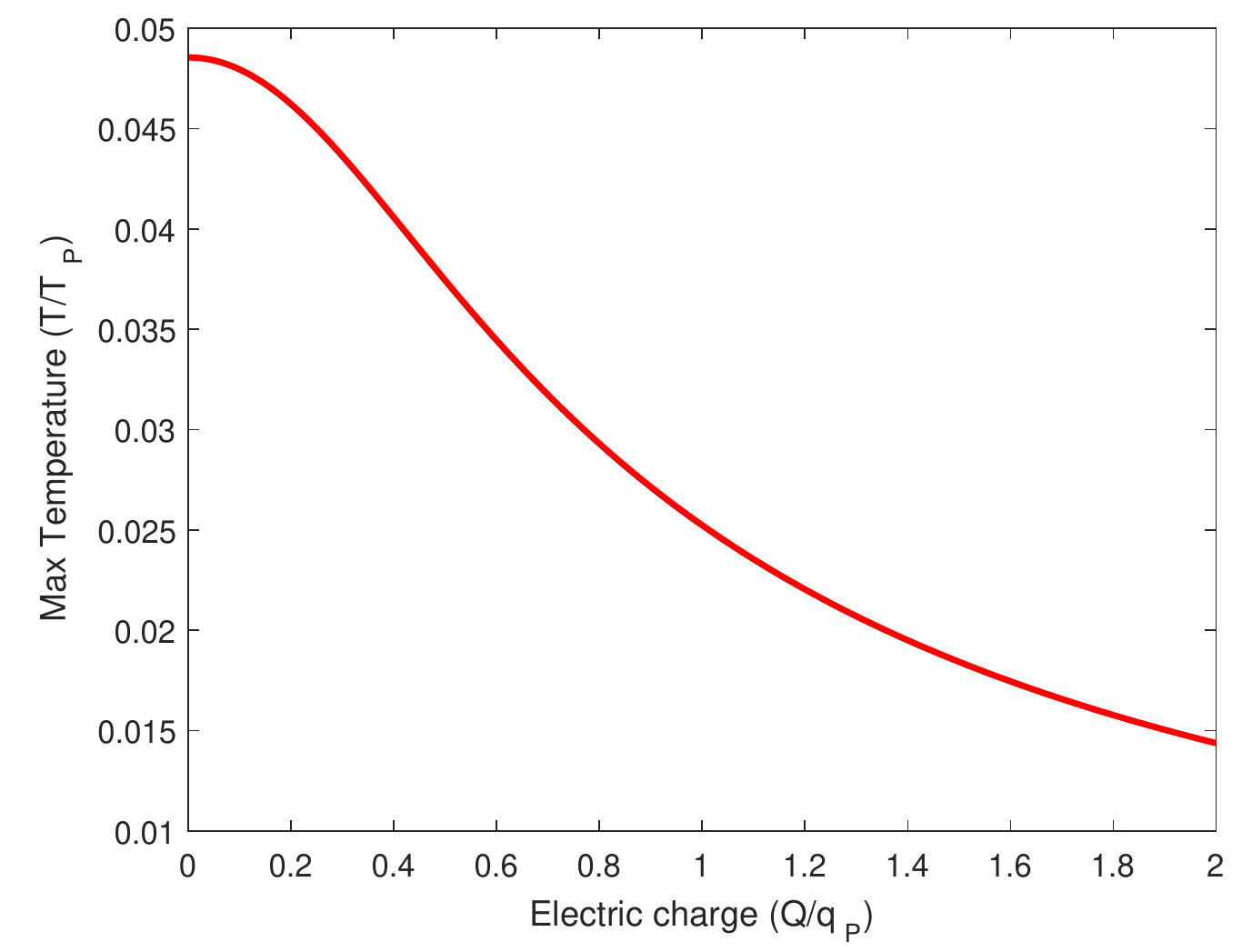}\par 
    \end{multicols}
\caption{Temperature of quantum black hole as a function of black hole mass and charge. Numerical values are $c_{A}=0.398$, and $c'_{A}=1.943$.\label{T_max_mass_charge}}
\end{figure}

The entropy of black hole obtains correction as well. For the $r=\tilde{r}_{+}$, the outer horizon, we have $f(\tilde{r}_{+},M,Q) = 0$. This equation, acting as the equation of state remains unchanged by varying the physical parameters. Hence we have,

\begin{equation}
\label{eq:partial}
\begin{split}
\left.\frac{\partial f}{\partial M }\right )_{\tilde{r}_{+},Q} \mathrm{d} M+\left.\frac{\partial f}{\partial Q}\right)_{\tilde{r}_{+},M} \mathrm{d} Q+\left.\frac{\partial f}{\partial \tilde{r}_{+}}\right)_{M,Q} \mathrm{d} \tilde{r}_{+}=0
\end{split}
\end{equation}

In terms of Hawking temperature, we can rewrite equation \eqref{eq:partial} in the standard thermodynamic form,

\begin{equation}
\label{eq:first law}
\begin{split}
dM = - \frac{\left. \frac{\partial f}{\partial Q}\right )_{r_+,M}}{\left. \frac{\partial f}{\partial M }\right )_{r_+,Q}} dQ - \frac{T_H}{\left. \frac{\partial f}{\partial M }\right )_{r_+,Q}} 4\pi d\tilde{r}_{+}
\end{split}
\end{equation}
Comparison with the standard thermodynamic relation $dE=TdS+\phi dQ$ gives the expression for $dS$ ,
\begin{equation}
\label{eq:entropy solving 1}
\begin{split}
dS = -\left(\frac{\partial f}{\partial M}\right)^{-1} 4\pi dr_+ = \frac{2 \pi r_+ dr_+}{ \left(1-c_A l_p^2 \left(\frac{2M}{r_+^3}-\frac{Q^2}{r_+^4}\right)\right)}
\end{split}
\end{equation}
Integrating this expression we obtain the corrected entropy, of course  to the first order of  $\hbar$, 

\begin{equation}
\label{eq:entropy}
\begin{split}
S = \frac{A}{4} + \pi c_A l_p^2 \ln(A)
\end{split}
\end{equation}
where the area $A$ is in terms of Planck units.
We draw the attention to the logarithmic correction of the entropy formula.
Note that $A$ is the corrected area which is $A_{RN}+O(l_{p}^{2})$.

From (\ref{eq:first law}), one can see that  the chemical potential $\mu$ takes the following form:

\begin{equation}
\mu=\frac{Q}{r_{+}} \left[1 - c_{A} l_{p}^{2} \left( \frac{2}{r_{+}^{2}} +  \frac{(1-2\alpha)Q^{2}}{r_{+}^4} \right)  \right]
\end{equation}

The chemical potential is defined to be the work needed to adiabatically increase the ADM mass, $M$, per unit of electric charge, $Q$. In the classical case the $\mu$ is given by $\frac{Q}{r_{+}}$ which is the same as electric potential. That is why the falling of a massless charged particle with charge $dq$ from $r=\infty$ to $r=r_{+}$ results in a $QdQ/r_{+}$ change in $M$ and no change in entropy, $S$.  After applying quantum correction, entropy gets modified and so the adiabatic paths get correction. Hence in order to keep the entropy unchanged under falling of a charged particle with electric charge $dQ$ and rest mass $dm$, $dm$ cannot be zero anymore. This nonzero rest mass results in a gravitational work appearing in the chemical potential.


We can obtain the heat capacity of the quantum corrected black hole at constant charge using (\ref{eq.4.1}),
\begin{align}
C_{Q} &=\left. \frac{\partial M}{\partial T_{H}} \right)_{Q} \nonumber\\
&=\frac{2 \pi M^{2} \eta_{+}^2 (\eta_{+}-1)}{3-2\eta_++\frac{2c_{A}l_P^2}{M^2\eta_+^3} \left(  -7\eta_+^2+19\eta_+-13+\frac{3\alpha}{4} (2-\eta_{+}) (2\eta_+^3-17\eta_+^2+\frac{104}{3}\eta_{+}-\frac{52}{3})\right)  }
\end{align}

A first order phase transition occurs where $C_{Q}$ diverge which can be shown, albeit after a long and tedious algebra, to take place at,
\begin{eqnarray}
\left. \left( \frac{Q}{M} \right) \right|_{\rm{critical}}=\frac{\sqrt{3}}{2}(1-\frac{c_Al_P^2}{M^2}(\frac{19\alpha-4}{81}))
\end{eqnarray}
\begin{eqnarray}\approx
0.866+\frac{c_{A}l_{p}^{2}} {M^{2}} \left( 0.043-0.235\alpha \right)
\end{eqnarray}
Giving corrections to criticl points of classical RN black hole. At this point the temperature of the black hole is maximum,
\begin{eqnarray}
T_{H}^{max}=\frac{1}{M}(0.035-\frac{c_{A} l_{p}^{2}}{M^2} (0.015+0.048\alpha)) \label{eq.4.10}
\end{eqnarray}
We shall note that these result is only valid for black holes with large charge and mass. In the next part  the special case with zero charge and small mass is considered.
\subsection{Thermodynamics of special case of Q=0\label{Thermodynamics of special case of Q=0}}

We saw in the previous section that for any fixed charge $Q$, the temperature vanishes at two limits of very large mass and at a finite mass which is strictly larger than the charge. Hence it must pass through a maximum. This phenomenon is more interesting when the charge goes to zero. We observe that the zero temperature zero charge state has a very special role. To see this, consider the third law of (black hole) thermodynamics and evaporation of the neutral black hole,  with quantum correction. Since this state has a finite mass that sets an absolute minimum to the mass of black holes, the evaporation can not go beyond it. This mass limit and its zero temperature show that the final stage of the evaporation of a black hole with zero charge, cannot be anything but this extremal state. The third law implies that this can not occur in finite steps or equivalently, this state can not be attained over a finite time. One can show that direct calculation of its lifetime results in the same conclusion.   This asymptotic state is the cold remnant of the black hole evaporation. In the path toward this point, the black hole must pass through a maximum temperature given by (\ref{eq.4.10}). To proceed further, we shall address this issue in a more direct way.

Let us take the black body radiation for black hole,
\begin{equation}
\label{eq:evaporation rate}
\begin{split}
-\frac{d M}{d t} = \sigma A_{H} T_{H}^4 = \sigma (4\pi r_+^2) \left( \left.\frac{\partial f}{\partial r}\right|_{r=r_+} \right)^{4} (4\pi)^{-4}
\end{split}
\end{equation}

 Using the above equation, we shall find lifetime of black hole $\tau$:

\begin{equation}
\label{eq:lifetime}
\begin{split}
\tau= -(4 \pi)^{-3} \sigma \int_{r_{+}^{0}}^{r_{ext}} \frac{d M}{d r_+} \frac{d r_+}{f^{\prime ^4}(r_+) r_+^2}
\end{split}
\end{equation}
where we put the radius of extremal black hole $r_{ext}$ as the limit of integral since it is where the temperature vanish and $r_{+}^{0}$ as the initial value of outer horizon. Note that $\partial M/\partial r_{+}$ and $r_{+}$ always have finite nonzero values. Let us consider this integral at the point of our interest, near extremal and Taylor expand the outer horizon around extremal point $r_{+}=r_{ext}+\epsilon$,

\begin{equation}
\label{eq:taylor epsilon}
\begin{split}
f'(r_{+}) = \frac{3}{2 c_A l_p^2} \epsilon -\frac{15}{4} \sqrt{\frac{3}{2}} \frac{1}{c^{\frac{3}{2}} l_p^3} \epsilon^2 + O(\epsilon^3)
\end{split}
\end{equation}
in which $r_{ext}=\sqrt{\frac{8}{3} c_A} l_p$ and $M=\sqrt{\frac{32}{27}c_{A}}l_{p}(1+\frac{21 \epsilon^{2}}{8c_{A}l_{p}^{2}})$.
Finally we obtain $\tau \sim -\int_{\epsilon^{0}}^{0} d\epsilon / \epsilon^{3} \sim \lim_{\epsilon \rightarrow 0} 1/\epsilon^{2}$ which is divergent. This shows the lifetime of the black hole, even after quantum correction is infinite.

\begin{figure}
\centering
\includegraphics[width=0.8\linewidth]{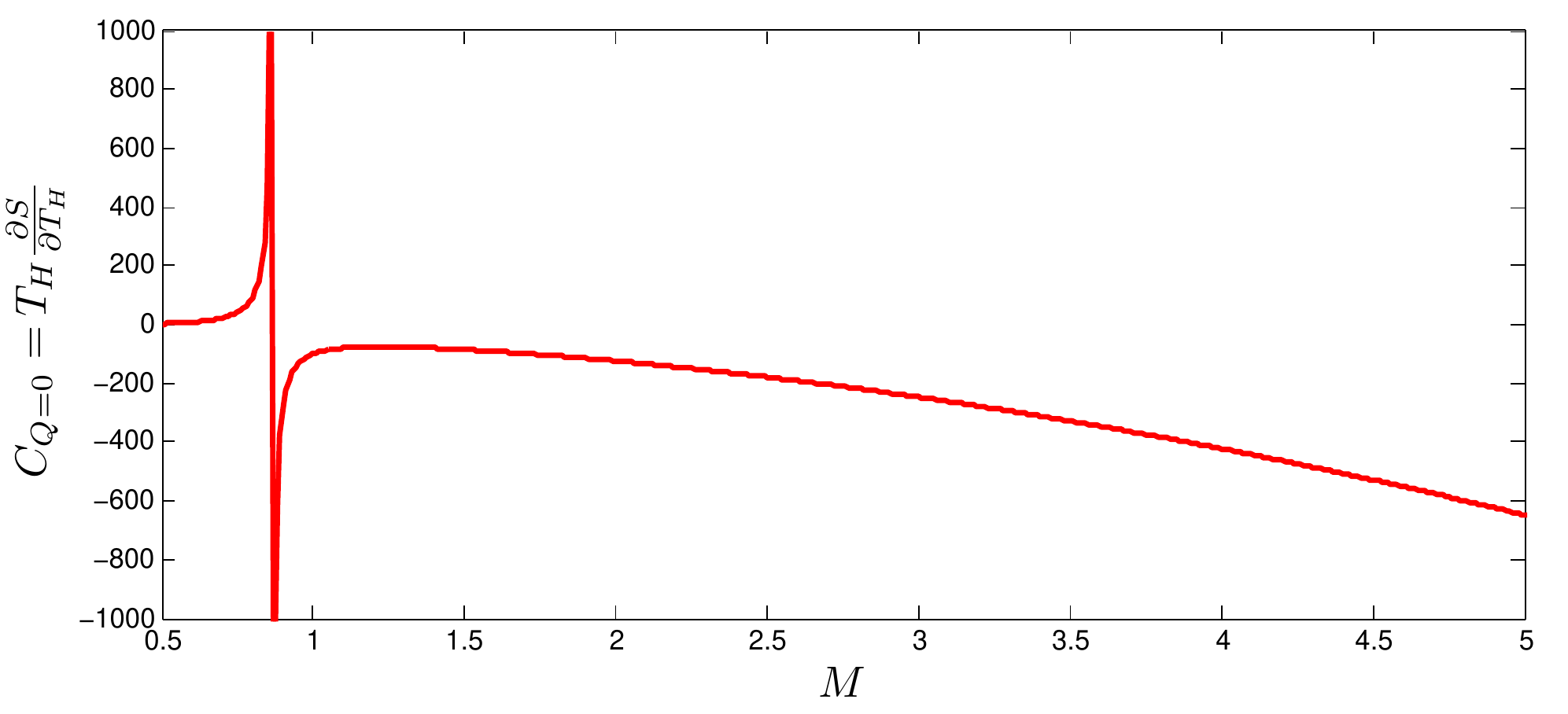}
\caption{Heat capacity for the special case ($Q=0$) as a function of mass of the black hole. Numerical value is $c_{A}l_{p}^{2}=1$}
\label{HeatCapacity}
\end{figure}

Let us now consider  the heat capacity ,
\begin{eqnarray}
C=T_{H}\frac{\partial S}{\partial T_{H}}=\frac{\partial M}{\partial T_{H}}=-\frac{8 \pi M^{2}}{1-\frac{3 c_{A}}{4} \frac{M_{p}^{2}}{M^{2}}} \label{eq.4.14}
\end{eqnarray}
As already mentioned we have a first order phase transition at the critical point where the temperature has a maximum and heat capacity  a jump ( figure \ref{HeatCapacity}).

It is worth remarking that in this calculation we obtain approximate critical point (see Appendix \ref{A2}), but the general behavior of the heat capacity is reflected in the (\ref{eq.4.14}). The reason is that there always exists a maximum temperature and there are two different states of charged black hole with the same temperature. For large black holes ($M>\sqrt{\frac{3}{4}c_{A}} M_{p}$) it is similar to the Schwarzschild black holes where the heat capacity is negative. The negative heat capacity signifies that a reduction in the system's energy (mass)  increases its temperature. This property is characteristic of systems with long-range attractive forces, and thus for systems with gravitational self-action. They include gravitational systems such as galaxies, stars, black holes, and also sometimes some nano-scale clusters of a few tens of atoms, close to a phase transition point \cite{Schmidt:2000zs}.

The most conspicuous feature is the infinite discontinuity at peak of the temperature, after which the specific heat changes from negative to positive. Therefore, $M=\sqrt{\frac{32}{9}c_{A}} M_{p}$ is the critical point, beyond which the heat capacity is positive with critical exponent 1.

The peak of the temperature which occurs at  $M=\sqrt{\frac{32}{9}c_{A}} M_{p}$ is,
\begin{eqnarray}
T_{H_{+}max}=\frac{1}{16 \pi \sqrt{2 c_{A}} M_{p}}
\end{eqnarray}
We note that the above  mass is slightly larger than the extremal mass of the Schwarzschild black hole.

At the critical point, the black hole enters a new phase in which new thermodynamic degrees of freedom appear.
Therefore as the temperature of the black hole grows, the degrees of the freedom raises and diverges at the critical point, where the temperature reaches the maximum value.

In the next section, we apply our results to two cases of a black hole in a volume in thermal equilibrium, and the collapse of a charged shell of dust.


\section{Applications\label{Applications}}
We consider two examples of t application of the quantum corrected metric. These examples are chosen to shed light on the effects of the quantum corrections by pointing to the difference with the standard RN or Schwarzschild cases. The first one is considerations concerning the equilibrium and stability of a black hole in a container. As we shall see the constraints on the container changes when the quantum effects become important i.e. 
when the black hole is small. The second application is the study of the collapse of a charged shell of dust which clarifies how the singularity is avoided even in the case of charge-less case.  

\subsection{Black hole in thermodynamic equilibrium}

One of the central achievements of thermodynamics is its ability to predict the conditions for stable equilibrium, even for multi-component systems,  using the second law. This treatment may also be applied to black holes in contact with heat reservoirs, energy sources, etc.
As an example, we consider a quantum black hole with mass $M$ residing inside an adiabatic enclosure with volume $V$, in equilibrium at a constant temperature. As expected the result for large black holes is the same as that obtained previously \cite{Davies:1977} with small quantum corrections.

If a black hole is placed in a container with perfectly reflecting walls, account must be taken of the accumulation of thermal evaporation radiation inside the container.
Under these circumstances, the equilibrium condition is that the total entropy of the contents of the container to be at an extremum. Stability occurs if the extremum is a maximum.
The total entropy of the black hole plus radiation is,
\begin{eqnarray}
S=S_{BH} + \frac{4}{3} \left( \alpha V M_{r}^{3} \right)^{\frac{1}{4}}
\end{eqnarray}

Where $M_{r}$ denotes the mass(energy) of the radiation, $V$ is the volume of the container and $\alpha$ is the radiation constant. The condition for equilibrium is $dS=0$. Subject to the adiabatic constraint and energy conservation $dM+dM_{r}=0$, the result is,
\begin{eqnarray}
\!\!\!\! \frac{\partial S}{\partial M}=\frac{1}{T_{H}}  -  ( \alpha V )^{\frac{1}{4}} M_{r}^{-\frac{1}{4}}=0
\end{eqnarray}
which gives,
\begin{eqnarray}
\alpha V= \frac{M_{r}}{ T_{H}^{4}}
\end{eqnarray}

For stability it is necessary to have $d^{2}S<0$. Differentiating dS a second time and using the obtained equation for $V$ yields,
\begin{eqnarray}
\frac{\partial^{2} S}{\partial M^{2}} = -\frac{1}{T_{H}^{2}} \frac{\partial T_{H}}{\partial M}  - \frac{1}{4} \frac{M_{r}^{-1}}{T_{H}} <0
\end{eqnarray}

Since there exist a maximum temperature $T_{H}^{max}$ at $M^{max}$ (which is at the order of Planck mass), there exist two sets of results $M>M^{max}$ and $M<M^{max}$.
First for the case $M>M^{max}$, which behaves similar to the Schwarzschild black hole, in this case $\partial T_{H}/\partial M <0$ and we obtain following constraint,
\begin{eqnarray}
M_{r} < -\frac{T_{H}}{4 \partial T_{H}/\partial M } 
\end{eqnarray}
This is an upper bound for mass of radiation within the container. For large mass taking $T_{H} \rightarrow 1/8\pi M$ it resembles the constraint of Schwarzschild black hole $M_{r}<M/4$. It also gives the constraint for volume of the container,
\begin{eqnarray}
\alpha V < -\frac{1}{4 T_{H}^{3} (\partial T_{H}/\partial M ) }
\end{eqnarray}
which approaches to the $\alpha V < \left(\frac{2\pi}{M_{p}^{2}}\right)^{4} M^{5}$ for large mass as expected is similar to the Schwarzschild black hole \cite{Davies:1977}.
We conclude that for the case $M>M^{max}$ it is not possible to set a black hole in equilibrium in empty space or a container with a volume larger than a certain amount. The black hole, in this case, is unstable and will radiate away its mass and shrinks down to the Planck length where the phase transition takes place.

For the second case $M<M^{max}$ which is of particular interest we obtain,
\begin{eqnarray}
M_{r} \geq 0, \ \ \ \ \ \ \ \ \alpha V \geq 0
\end{eqnarray}
This shows that after the phase transition the quantum black hole enters into a new regime where it is stable and stays in equilibrium with the environment without imposing any constraint.

\subsection{Collapsing charged shell}

In this part, we study the gravitational collapse of a spherical and charged shell of dust with the quantum corrections to the background geometry taken into account. Assume the 3-manifold, $\Sigma$, to be the trace of the shell as it moves in the space-time. As $\Sigma$ splits space-time into two different regions, $V^-$, inside the shell and $V^+$, out of it.
During the collapse, the geometry stays spherically symmetric due to the initial condition. Since neither gravitational nor electromagnetic radiation propagates as S-waves, the geometry stays static in $V^{\pm}$.
This shows that the solution outside and inside the $\Sigma$ should be time independent except for the movement of the shell. Hence, the static solution in $V^+$ must be the unique spherically charged solution we obtained in section \ref{Geometric modifications} and the static inner solution should remain flat, same as it is in the infinite past where the shell is located in infinity. The dynamics of the radius is dictated by the matching condition of the inner and outer metrics.

We extend the coordinates $r, \theta, \phi$ defined on $V^+$ to the inner region. Since the metric inside is flat, one can find a time coordinate $t_-$ so that the metric in $V^{-}$ takes the following form,
\begin{align}
ds^2=-dt_-^2+dr^2+r^2d\Omega^2
\end{align}

Before further steps, we should make sure as part of matching conditions, that the line element changes continuously as we cross the shell. Note that the interior and exterior coordinates do not necessarily join continuously ($t_{+} \neq t_{-}$ on $\Sigma$). To be more precise, we should check if both metrics induce the same metric on the hypersurface $\Sigma$. This matching condition has been known for a long time \cite{Israel:1966}. We parametrize $\Sigma$ by $\tau, \theta$ and $\phi$, which $\tau$ is the proper time of a radially infalling observer pinned to the shell.

The vector $\partial_{\tau}$ is a linear combination of $\partial_{r}$ and $\partial_{t}$ and is independent from $\theta$ and $\phi$. Hence we conclude:
\begin{itemize}
\item The induced metric, $g^s_{ab}$, should be diagonal in terms of $\tau, \theta$ and $\phi$
\item $g^s_{\theta\theta}=g^+_{\theta\theta}$ and $g^s_{\phi\phi}=g^+_{\phi\phi}$.
\end{itemize}
here $\pm$ and $s$ indicate the regions $V^{\pm}$ and $\Sigma$ respectively.

Additionally, $\tau$ is set to be the proper time of an observer moving on $\Sigma$ with $d\theta=d\phi=0$ which means $g^s_{\tau\tau}=-1$. Therefore, the induced metric in terms of the coordinates on the shell takes the following form.
\begin{align}
ds^{2}=- d\tau^{2} + R(\tau)^{2}(d\theta^{2}+ \sin^{2}\theta d\phi^{2})
\end{align}
where $R(\tau)$ stands for the shell's radius at a specific proper time $\tau$. Our goal is to find the evolution of the shell's radius, $R$, in terms of its proper time, $\tau$. Matching the metrics induced from outside and inside of the $\Sigma$ we have:
\begin{align}
dt_-^2-dr^2=d\tau^2=f(R)dt^2-f(R)^{-1}dr^2
\end{align}
or
\begin{align}
f(R)^{2}(dt/d\tau)^{2}-f(R)=(dR/d\tau)^{2}=(dt_-/d\tau)^{2}-1
\end{align}
Using this matching conditions, the extrinsic curvatures on $V^{\pm}$ \cite{poisson} are
\begin{eqnarray}
K_{\pm}^{\tau}{}_{\tau}=\dot{\beta_{\pm}}/\dot{R},
\end{eqnarray}
\begin{eqnarray}
K_{\pm}^{\theta}{}_{\theta}=K_{\pm}^{\phi}{}_{\phi}=\beta_{\pm}/R,
\end{eqnarray}
where
\begin{eqnarray}
\beta_{+}=\sqrt{\dot{R}^{2}+1-\frac{2M}{R} + \frac{Q^{2}}{R^{2}} + c_{A} l_{p}^{2} \left( \frac{2 M^{2}}{R^{4}} - \frac{2MQ^{2}}{R^{5}} +\frac{\alpha Q^{4}}{R^{6}} \right) }
\end{eqnarray}
\begin{eqnarray}
\beta_{-}=\sqrt{\dot{R}^{2}+1}
\end{eqnarray}

\begin{figure}[H]
\centering
\includegraphics[width=\linewidth]{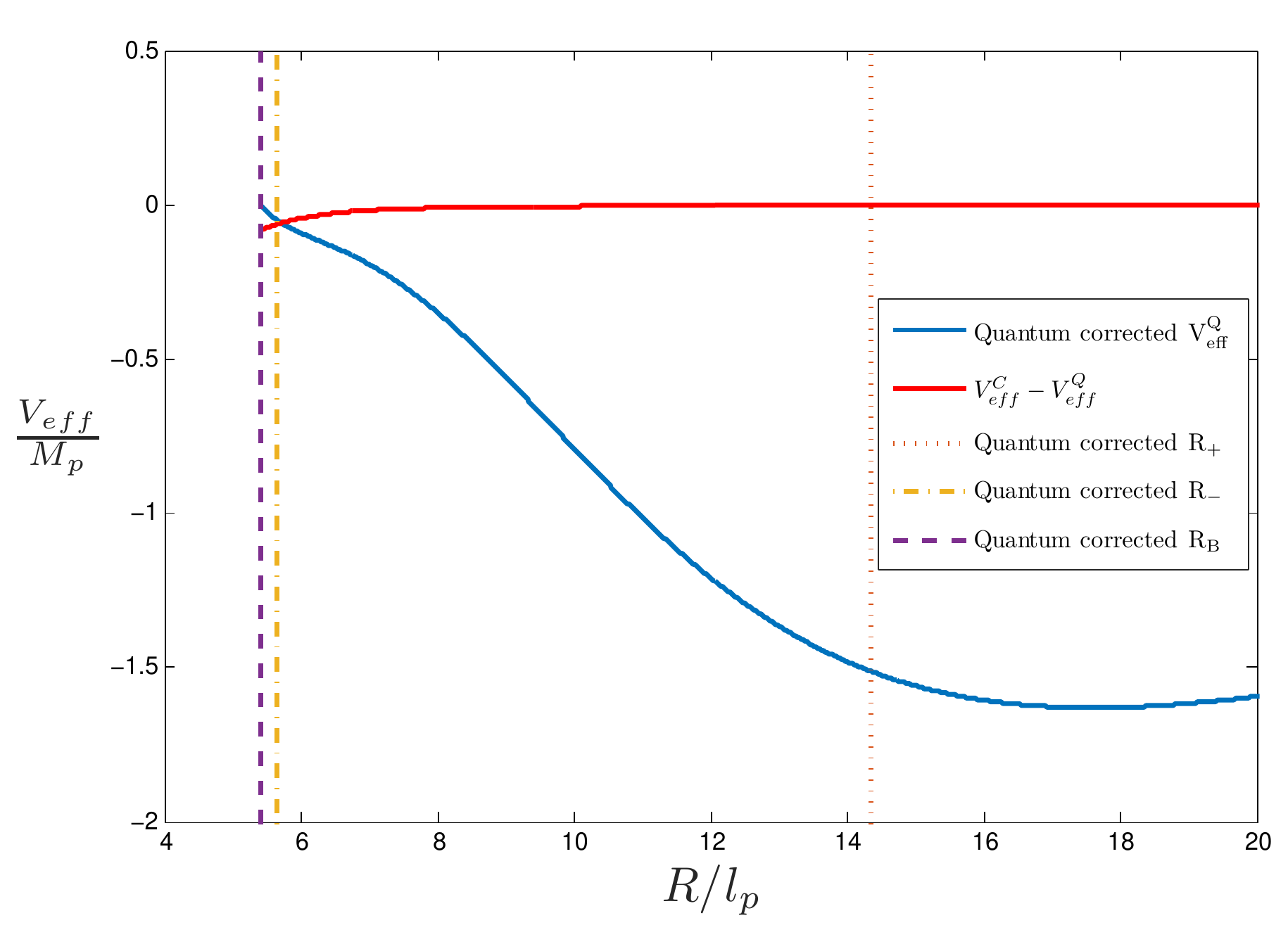}
\caption{Corrected potential $V^Q_{eff}$ (blue), the amount of correction $V^C_{eff}-V^Q_{eff}$ (red), quantum corrected horizons (orange and yellow) and bounce radius $R_B$ (purple) for numerical values $(m,Q)=(10m_P,9q_P)$.}
\label{ShellCollapse}
\end{figure}

The Einstein equations gives that the surface stress-energy tensor $S_{ab}$ should be equal to the following purely geometrical and diagonal tensor defined on $\Sigma$\cite{poisson}.
\begin{align}
S_{ab}&=-\frac{1}{8\pi}([K_{ab}]-[K]h_{ab})\nonumber\\
S^{\tau}_{\tau}&=\frac{1}{4\pi R}(\beta_{+}-\beta_{-})\nonumber\\
S^{\theta}_{\theta}&=S^\phi_\phi=\frac{1}{8\pi R\dot R}\frac{d}{d\tau}(R(\beta_{+}-\beta_{-}))
\label{SEMT}
\end{align}
where $h_{ab} = g_{ab} + u_{a}u_{b}$ that $u_{\mu}$ stands for the velocity of the shell, satisfying $u_{\mu}u^{\mu}=-1$. Our next step would be to find and substitute the surface energy momentum tensor in the equation \eqref{SEMT}. Diffrentiating the non gravitational part of the Lagrangian with respect to the metric gives the following energy momentum tensor on $\Sigma$. Note that we have expressed the energy momentum tensor in the coordinate system defind on $V^+$ i.e. $\mu\in\{t,r,\theta,\phi\}$.
\begin{align}
T^{\mu\nu}=\bigg [\sigma(\tau) u^{\mu}u^{\nu}+\frac{Q^2}{4\pi R^3u^{t}}(\delta^\mu_{t} u^\nu+\delta^\nu_{t} u^\mu-g^{\mu\nu}g_{tt}u^t)\bigg]\delta(r-R(\tau))+T_F^{\mu\nu}
\end{align} 
The $T_F^{\mu\nu}$ is a finite part and does not affect the surface stress-energy tensor. The surface stress-energy tensor is defined as $S_{ab}=t_{\mu\nu}e_a^\mu e_b^\nu$ where $\{e_a^\mu\}_{a\in\{\tau,\theta,\phi\}}^{\mu\in\{t,r,\theta,\phi\}}$ represent the transform coefficients i.e. $\partial_a=e_a^\mu\partial_\mu$ and $t_{\mu\nu}\delta(r-R(\tau))$ is the infinite part of $T_{\mu\nu}$ on $\Sigma$. Based on the definitions of coordinate systems we have,
\begin{align}
e_\tau^\mu&=u^\mu\nonumber\\
e_\theta^\mu&=\delta^\mu_\theta\nonumber\\
e_\phi^\mu&=\delta^\mu_\phi
\end{align}
Therefore the surface energy-momentum tensor, $S^a_b$ takes the following form.
\begin{align}
S^\tau_\tau&=-\sigma+\frac{3Q^2 f(R)}{4\pi R^3}\nonumber\\
S^\theta_\theta&=S^\phi_\phi=\frac{Q^2 f(R)}{4\pi R^3}
\end{align}
By substituting $S$ in the equation \eqref{SEMT}, we have
\begin{align}
-4\pi R^2\sigma+\frac{3Q^2 f(R)}{ R}&=R(\beta_{+}-\beta_{-})\label{SC1}\\
\rightarrow \frac{2Q^2f(R)\dot R}{ R^2}&=\frac{d}{d\tau}(R(\beta_{+}-\beta_{-}))\label{SC2}
\end{align}
Integrating the above equation gives,
\begin{align}
&\frac{d}{d\tau}(4\pi R^2\sigma-\frac{3Q^2 f(R)}{ R}+2Q^2\hat{f}(R))=0\nonumber\\
\rightarrow &4\pi R^2\sigma-\frac{3Q^2 f(R)}{ R}+2Q^2\hat{f}(R)=m:cte \label{eq.5.22}
\end{align}
where $m$, a constant of motion, is the shell's rest mass. $F(R)$ satisfies $\frac{d}{dR}
\hat{f}(R)=\frac{f(R)}{R^2}$ and is given as
\begin{align}
\hat{f}(R)=\frac{1}{R}-\frac{M}{R^2}+\frac{Q^2}{3R^3}+c_Al_p^2\left(\frac{2M^2}{5R^5}-\frac{MQ^2}{3R^6}+\frac{\alpha Q^4}{7R^7}\right)
\end{align}
The Eq. (\ref{eq.5.22}) gives $\sigma$ in terms of $R, Q$ and $m$ which transforms equation \eqref{SC1} to,
\begin{eqnarray}
\dot{R}^{2}+V_{eff}(R)=0,\label{eq2.36}
\end{eqnarray}
where
\begin{align}
V_{eff}(R)&=1-\frac{(f(R)-1-U(R))^2}{4U(R)};\nonumber\\
U(R)&:=\frac{(m-2Q^2\hat{f}(R))^{2}}{R^{2}}
\end{align}
To find the bounce radios we set $\dot{R}^{2}=-V_{eff}(R)=0$. It gives the corrected bounce radius $R_{B} $ which is inside the inner horizon $ \tilde{R}_{-}$. Fig. \ref{ShellCollapse} shows $V_{eff}$ for a dust shell as a function of $R$.
As one can see in figure \ref{ShellCollapse}, the plot of $V_{eff}$ is shifted up after taking quantum corrections into account. Therefore quantum corrections increases the bounce radius.
\section{Conclusion}
For a better understanding of  back reaction  of QFT fluctuations  on the background geometry, we have  extended our previous results on Schwarzschild solutions \cite{Abedi:2015yga,Arfaei:2016dbh} to the Reissener Nordstrom case. The corrections provide significant changes to both geometry and thermodynamics of black holes, charged or uncharged, especially when it becomes small, i.e. approaching Planck scales. Our approach in this work is semi-classical:  the gravity is considered classical and the fields living in the gravitational background as quantum. Since we have looked for static and spherically symmetric solutions, the knowledge of the trace of the energy momentum tensor induce by the quantum fluctuations, has been sufficient to find all the necessary elements for our purpose.  The changes to both internal and external horizons  turn out to be  small for macroscopic ones, this change takes place in the same region that other quantum gravity effects arise \cite{THOOFT1985727,Abedi:2016hgu}. Although the changes are small, they have significant effects to the extent of stopping the process of collapse before reaching the singularity. The main reason for this phenomenon is existence  of the  internal horizon in which the radial direction becomes a space-like allowing avoidance of the singularity by the in-falling matter. This is an unexpected phenomenon for the $Q=0$ case. Another significant result is the change in the curve that separates the region of the solutions with naked singularity (considered unphysical ) from the acceptable solutions, the extremal curve. We get approximate expressions for this curve for two regions; large Q and M, and very  small charge and mass close to Planck scales, more accurately for two regimes of $QM> c_A l_p^2$  and $QM<  c_A l_p^2$. The curve is calculated by numerical techniques and is shown in Fig. \ref{Temperature}.  For a given charge, we find a lower bound for the mass which extends to a non-zero value  at zero charge. The lower bound mass specifies an extremal solution with double horizons, again even for the zero charge limit. We will get back to this subject at the end of this section.
It is shown in the appendix \ref{A1} that the number of horizons remains unchanged for the charged black hole which is inherited to the Schwarzschild case. The physical reason behind this phenomenon lies in the fact that the energy-momentum tensor of a point charge and the anomaly induced energy-momentum tensor both behave in the same manner. The responses of the gravitational field are similar in both cases.

It is clear that our approximation is not accurate for very small black holes but one can trust the general picture that is emerging for small black holes, such as the existence of an extremal zero temperature Schwarzschild Black holes with a mass of order $M_{p}$.

It is well-known that the RN black hole attains a maximum temperature as mass decreases. We find a correction to this maximum temperature which becomes an absolute maximum for zero charge. This temperature signifies a phase transition from negative specific heat to positive values. It is comparable to Hagedorn temperature in string theory if the string scale is taken to be the same as Planck scale \cite{Atick:1988}. String states turn into black holes once their mass exceeds Hagedorn temperature. Apart from the counting of black holes, this may provide another point of contact with string theory.

As discussed on section \ref{sec: diagonal T} existence of $r^{-4}$ term in the trace anomaly may appear in the off-diagonal terms, making the black hole dynamical. This effect comes from $F^{2}$ term in anomaly considering the pssibility of an external gauge field $F_{\mu\nu}$ which is discussed in \cite{Duff1977334} and is under investigation which will be reported elsewhere.

The correction to thermal properties is remarkable, apart from setting an absolute to temperature, the entropy receives logarithmic correction. Although the Electromagnetic potential does not change, the chemical potential changes to accounting for the corrections to the geometry. Despite the changes to the geometry and thermodynamics, we can not make new comments on the information problem. The extremal black holes, remnants under Hawking radiation, are of the order of Planck scales and cannot carry the amount of information to resolve the information problem.

Real black holes certainly carry angular momentum, thus it is of utmost importance to find out what happens in the case of rotating black holes. 
We are studying the case, but it has proved to be more challenging than the cases of spherically symmetric. We expect rotating black holes to emit their angular momentum through Hawking radiation and end up with the same final state, the extremal Schwarzschild black holes.

Our consideration has strong implications for the remnant of black holes for both charged and uncharged ones. It is believed that charged black holes lose their charge very quickly and turn into Schwarzschild black holes\cite{Abedi:2015yga}. But due to the small temperature and height of angular momentum barriers that become comparable to the Planck mass, even for the case of S-waves, the final remnant of charged black holes may become meta-stable, which is stable in the case of vanishing charge. Since we have found that there is a minimum mass of the order of Planck mass ($\sqrt{\frac{32c_A}{27}}M_{p}$ in our approximation) with zero charge, it shall be absolutely stable, with no interaction except gravitational. \bf These stable extremal black holes with only gravitational interaction provide a reasonable candidate for dark matter. \normalfont The negative pressure due to quantum fluctuations of fields from outside pulls its double horizon and keeps it stable against the gravitational attraction trying to compress it towards the singularity. If we assume all dark matter is made of such extremal black holes, their density should be approximately one in $10^{13}$ cube meter. In other words, there can exist about $10^{7}$ in the whole of the earth. We are investigating this possibility and will report our findings shortly.

\appendix
\section{Number of horizons\label{A1}}
In this appendix, our aim is to prove that the maximum number of horizons remains unchanged under quantum correction. The first order modified metric can be written as,

\begin{equation}
\label{eq:metric}
{ds}^2=P(r)r^{-6}{dt}^2-{P(r)^{-1}}r^6{dr}^2-r^2(d{\theta{}}^2+\sin^2{\theta{}}\
d{\phi{}}^2)
\end{equation}
where

\begin{align}
\label{eq:rhoandp:2}
P\left(r\right)\coloneqq r^6-2Mr^5+Q^2r^4+2c_Al_p^2M^2r^2-2c_Al_p^2MQ^2r+c_Al_p^2Q^4\frac{\left(3+\rho{}\right)}{5}
\end{align}
and
\begin{align}
\rho(r)\coloneqq\frac{c_A^{\prime}}{2c_A}
\end{align}
In order to find the horizons, one has to solve for the roots of the sixth degree polynomial $P(r)$. By considering $Q$ to be zero, the modified metric \eqref{eq:metric} consistently reduces to the modification of the Schwarzschild solution obtained in \cite{Abedi:2015yga}. There, it was shown that at most two horizons would develop. Here we go through a similar procedure to show that the maximum number of horizons will not exceed two. In other words, $P(r)$ can have at most two real solutions.

We break $P(r)$ into sum of two simpler polynomials:

\begin{equation}
\label{eq:solving2}
\begin{split}
P\left(r\right)=P_1\left(r\right) - P_2\left(r\right)
\end{split}
\end{equation}
where,
\begin{equation}
\label{eq:solving2}
\begin{split}
P_1\left(r\right) \coloneqq r^4\left(r^2-2Mr+Q^2\right)\ \ ,\
P_2\left(r\right)\coloneqq - c_Al_p^2(2M^2r^2-2MQ^2r+Q^4\frac{\left(3+\alpha\right)}{5})
\end{split}
\end{equation}

\begin{figure}
\centering
\includegraphics[width=0.7\linewidth]{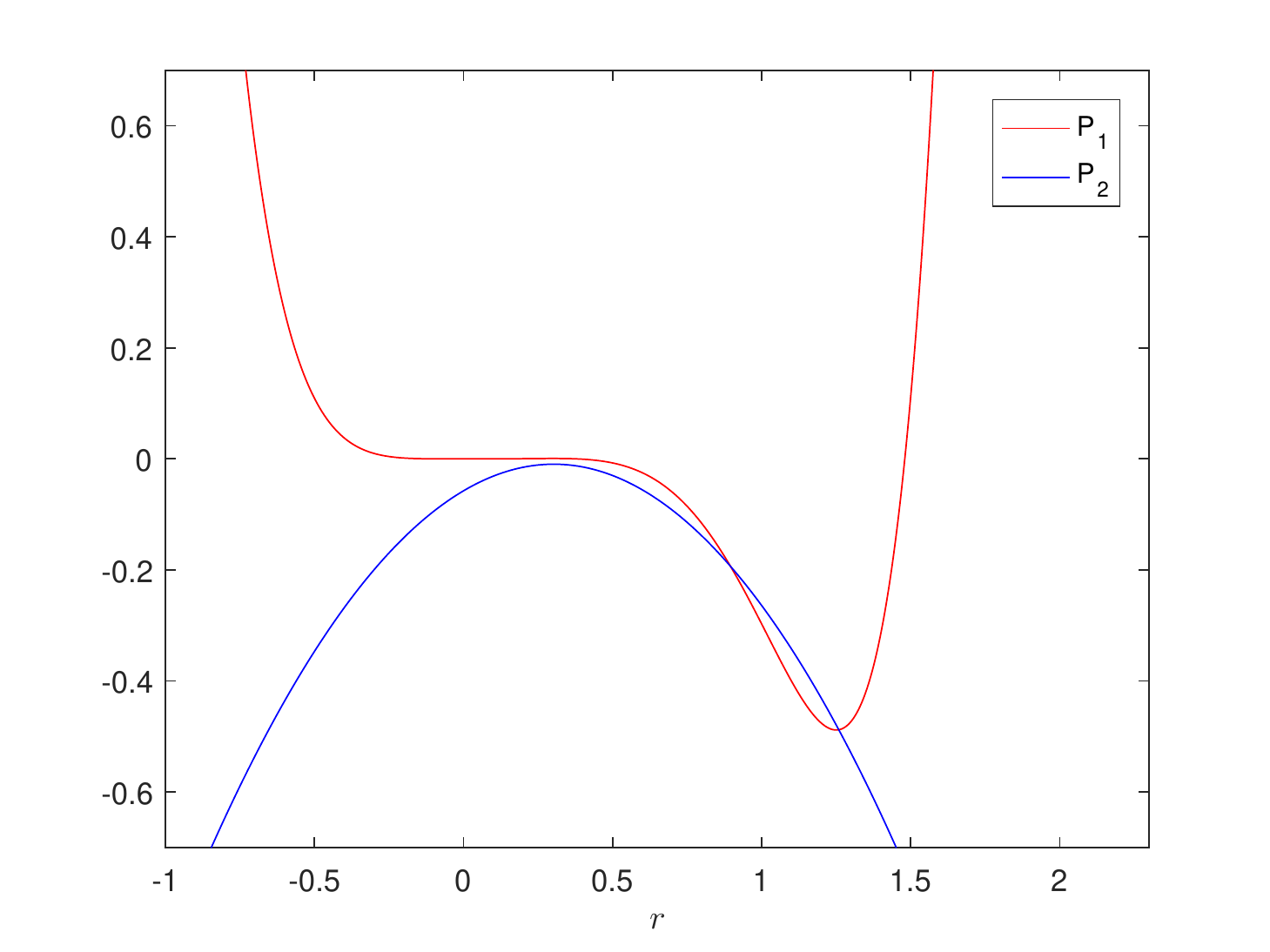}
\caption{Plots of $P_1(r)$ and $P_2(r)$ for $M=0.93$, $Q=0.75$, $c_A=0.303$ and $c_A'=0$.}
\label{polynom}
\end{figure}

The polynomial  $P_1$ comes from the Reissner-Nordstrom metric and its roots point to the classical horizons, while the polynomial $P_2$ stands for the quantum corrections which is expected to correct the roots of $\ P_1$. The polynomial $P_1(r)$ has four degenerate roots at zero and two other roots equal to $M\pm\sqrt{M^2-Q^2}$. The polynomial $P_2(r)$ is a second degree polynomial with negative discriminant ($\Delta_{P_2}=-4M^2Q^4\frac{1+\alpha}{5}<0$). So $P_2$ is a strictly negative polynomial which attains its maximum at $Q^2/2M$ (figure \ref{polynom}). Henceforth, $P(r)=P_1(r)-P_2(r)$ is definitely positive valued along real numbers except for the interval $I=(M-\sqrt{M^2-Q^2},M+\sqrt{M^2-Q^2})$ which has to be carefully studied, because $P_1$ and $P_2$ are both negative over there.
One can easily check that $\frac{d^2P_1}{dr^2}$ has three distinct real roots equal to $\{0,\frac{2}{3}M\pm\frac{2}{3}\sqrt{M^2-\frac{9}{10}Q^2}\}$. We show that at most one of them belongs to $I$. Obviously $0$ does not belong to $I$ and for proving our claim, it would be sufficient to show that $\frac{2}{3}M-\frac{2}{3}\sqrt{M^2-\frac{9}{10}Q^2}$  is not large enough to place in $I$. So it would be enough to prove the following inequality:
\begin{align}
\label{convexineq}
\frac{2}{3}M-\frac{2}{3}\sqrt{M^2-\frac{9}{10}Q^2}<M-\sqrt{M^2-Q^2}\leftrightarrow \sqrt{M^2-Q^2}-\frac{2}{3}\sqrt{M^2-\frac{9}{10}Q^2}<\frac{M}{3}
\end{align}
We study this inequality in two cases:

Case 1, ($|\frac{Q}{M}|>\sqrt{\frac{25}{27}}$):
\begin{align}
|\frac{Q}{M}|>\sqrt{\frac{25}{27}}=\sqrt{\frac{50}{54}}\rightarrow \frac{6}{10}Q^2>\frac{5}{9}M^2\rightarrow \frac{4}{9}M^2-\frac{4}{10}Q^2>M^2-Q^2\rightarrow\nonumber\\
\frac{2}{3}\sqrt{M^2-\frac{9}{10}Q^2}>\sqrt{M^2-Q^2}\rightarrow \sqrt{M^2-Q^2}-\frac{2}{3}\sqrt{M^2-\frac{9}{10}Q^2}<0<\frac{M}{3}
\end{align}
Case 2, ($|\frac{Q}{M}|<\sqrt{\frac{80}{81}}$):
\begin{align}
&|\frac{Q}{M}|<\sqrt{\frac{80}{81}}\rightarrow\frac{81}{400}Q^4<\frac{2}{10}M^2Q^2\rightarrow\nonumber\\
&M^4+\frac{441}{400}Q^4-\frac{21}{10}M^2Q^2<M^4+\frac{9}{10}Q^4-\frac{19}{10}M^2Q^2\rightarrow (M^2-\frac{21}{20}Q^2)^2<(M^2-Q^2)(M^2-\frac{9}{10}Q^2)\nonumber\\\rightarrow
&M^2-\frac{21}{20}Q^2<\sqrt{(M^2-Q^2)(M^2-\frac{9}{10}Q^2)}\rightarrow
\frac{12}{9}M^2-\frac{14}{10}Q^2<\frac{4}{3}\sqrt{(M^2-Q^2)(M^2-\frac{9}{10}Q^2)}\rightarrow\nonumber\\
&\frac{12}{9}M^2-\frac{14}{10}Q^2-\frac{4}{3}\sqrt{(M^2-Q^2)(M^2-\frac{9}{10}Q^2)}<0\rightarrow
(\sqrt{M^2-Q^2}-\frac{2}{3}\sqrt{M^2-\frac{9}{10}Q^2})^2-(\frac{M}{3})^2<0\nonumber\\\rightarrow& \sqrt{M^2-Q^2}-\frac{2}{3}\sqrt{M^2-\frac{9}{10}Q^2}<\frac{M}{3}
\end{align}
So the inequality \eqref{convexineq} holds either for $|\frac{Q}{M}|>\sqrt\frac{25}{27}$ or $|\frac{Q}{M}|<\sqrt{\frac{80}{81}}$ . So we proved the inequality for every amount of $|\frac{Q}{M}|$ and the proof of our claim is completed, therefore $\frac{d^2P_1}{dr^2}$ has at most one root placed in $I$. 

Obviously each intersection of the curves related to $P_1$ and $P_2$ corresponds to a real root of $P=P_1-P_2$ which results in a horizon (figure \ref{polynom}). The polynomial $P_2$ is a parabolic and hence a convex curve. So $P_1$ and $P_2$  will have at most two intersections if $\frac{d^2P_1}{dr^2}$ does not change sign within $I$. Let us consider the only chance to find four or more intersections i.e. the second derivative of $P_1$ changes its sign and finds a root in $I$. Since we proved that $\frac{d^2P_1}{dr^2}$ has at least two (of all its three) real roots placed before $I$, by checking the sign of $\frac{d^2P_1}{dr^2}$ one can conclude that $P_1$ has to be concave at the point $x=M-\sqrt{M^2-Q^2}$ in this case. Note that that at this point $P_1(x)=0>P_2(x)$ which makes it impossible to find four real roots by a single change of convexity over $I$. So polynomial $P$ has at most two real roots and modified Reissner Nordstrom black hole has at most two horizons.

\section{A note on non-analyticity\label{A2}}
\begin{figure}
\centering
\includegraphics[width=0.7\linewidth]{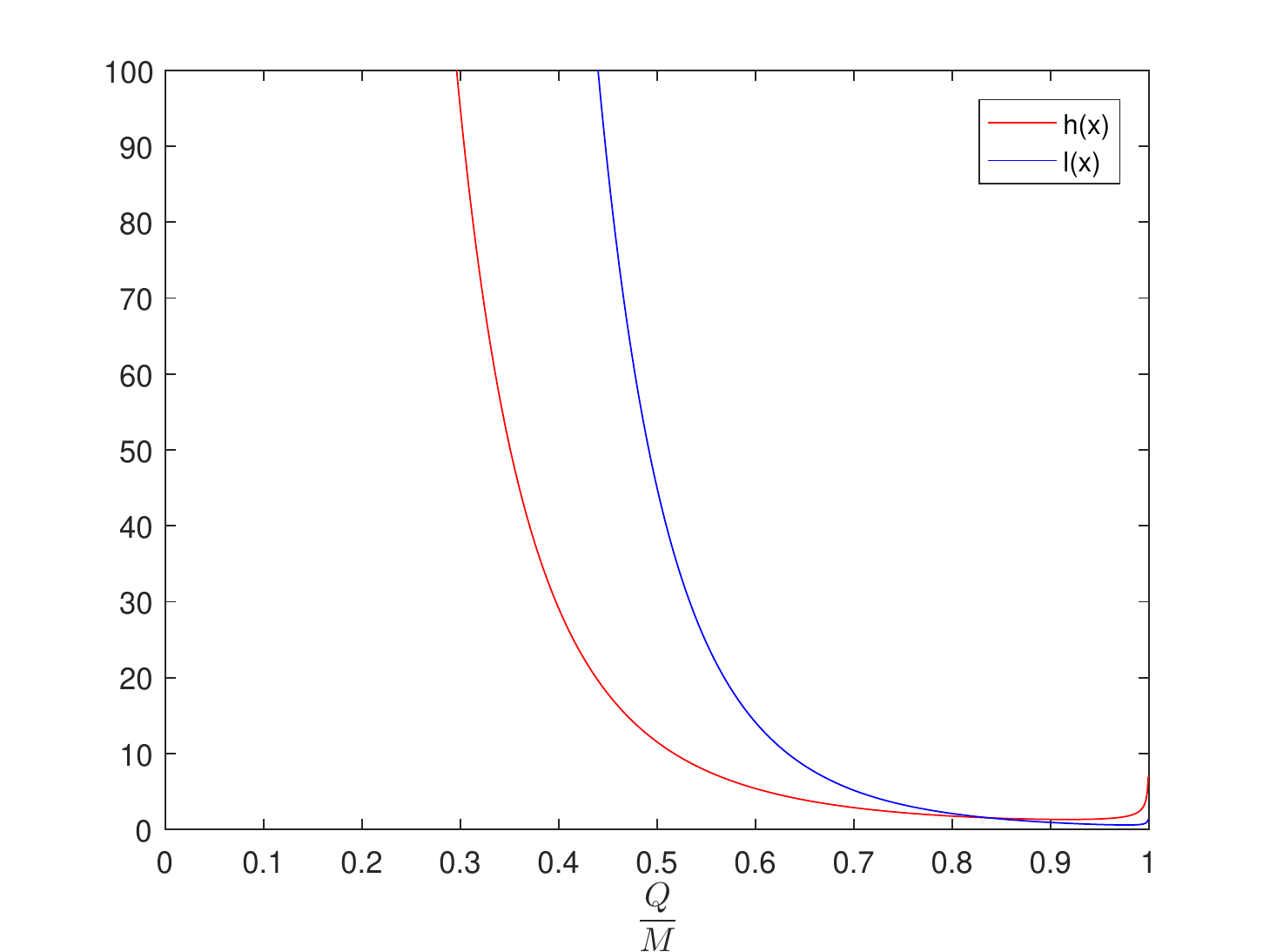}
\caption{$h(x)$ and $l(x)$, the coefficients appearing in the first order expansion of $\tilde r_-$.}
\label{nonanalyticity}
\end{figure}

Consider $c$ and $c'$ to be respectively defined as $\sqrt{c_A}l_P$ and $\sqrt{c_A'}l_p$. The first order expansion of corrected inner horizon ($\tilde r_-$) in terms of $c/M$ and $c'/M$ takes the following form:
\begin{align}
\label{firstorderroot}
\tilde r_-\approx M-\sqrt{M^2-Q^2}+h(\frac{Q}{M})\frac{c^2}{M}+l(\frac{Q}{M})\frac{c'^2}{M}
\end{align}
where,
\begin{subequations}
\label{eq:f and g}
\begin{align}
\label{eq:f and g:f}
h(x):=\frac{-2{\left(1-\sqrt{1-x^2}\right)}^2-2x^2\left(1-\sqrt{1-x^2}\right)+\frac{3}{5}x^4}{6{\left(1-\sqrt{1-x^2}\right)}^5-10{\left(1-\sqrt{1-x^2}\right)}^4+4x^2{\left(1-\sqrt{1^2-x^2}\right)}^3}
\\
\label{eq:f and g:g}
l(x):=\frac{-\frac{x^2}{10}}{6{\left(1-\sqrt{1-x^2}\right)}^5-10{\left(1-\sqrt{1-x^2}\right)}^4+4x^2{\left(1-\sqrt{1^2-x^2}\right)}^3}
\end{align}
\end{subequations}
As one can see in figure \ref{nonanalyticity},  $h(x)$ and $l(x)$ diverge as $x=\frac{Q}{M}$ tends to zero. This is physically meaningless since according to the equation \ref{firstorderroot}, it points out that the smaller horizon goes to infinity! In order to find that what is going on, let us take a look at the smaller horizon in the case of $Q=0$.

In this case the Q-dependent terms of  metric vanishes and it reduces to the following form,
\begin{equation}
\label{eq:solving1}
{ds}^2=P(r)r^{-6}{dt}^2-{P(r)^{-1}}r^6{dr}^2-r^2(d{\theta{}}^2+\sin^2{\theta{}}\
d{\phi{}}^2)
\end{equation}
where,
\begin{align}
P(r)=r^6-2Mr^5+2c^2M^2r^2
\end{align}
The smaller horizon $\tilde r_-$ is small, comparing to $M$, so that ${\tilde r_-}^6$ is negligible with respect to $2M{\tilde r}^5$. So we have,
\begin{align}
0\approx P(\tilde r_-)\approx {\tilde r_-}^6+2c^2M^2{\tilde r_-}^2\rightarrow \tilde r_-\approx M^\frac{1}{3}c^\frac{2}{3}
\end{align}
Hence, for $Q=0$, by decreasing quantum corrections ($c$),  $\tilde r_-$ tends to the classical smaller root ($r_- =0$) with rate of $c^\frac{2}{3}$ while this rate is of $O(c^2)$ for every nonzero $Q$. In fact this is because the normalized inner horizon, $\tilde r_-/M$, is a non analytic function in terms of $Q/M$, $c/M$ and $c'/M$. However it might, and probably is, analytic in terms of $c/M$ and $c'/M$ for a given $Q/M$. So for smaller amounts of $Q/M$, in order to get a better approximation for $\tilde r_-$, we should go through further perturbation orders. Or in other language, the first order perturbation is reliable for smaller amounts of $c/M$ and $c'/M$ (for larger masses) as we move to small $Q/M$ region. This is quite natural, since $Q/M\ll c/M$ is a different perturbative region. In fact, we showed that the inner horizon is not analytic in terms of Q, M, $c$ and $c'$, if one views $c$ and $c'$ as calculational parameters free to vary. But as is known, for fifth, or higher, degree polynomials there is no algebraic formula for roots in terms of polynomials coefficients. So the inner horizon can be non-analytic while the exact solution is so.

\acknowledgments

The authors would like to thank N. Afshordi, F.Ardalan, R. Mansouri and J. Tagizadeh Firouzjaei and especially Ali Naseh for useful conversations. We especially thank the anonymous referee for the valuable comments and suggestions to improve the clarity and the quality of this paper.


\end{document}